\documentclass[authoryear,preprint,review,12pt]{elsarticle} 
\usepackage[utf8]{inputenc}
\usepackage{xcolor}
\usepackage{todonotes}
\usepackage{psfrag}
\usepackage{ulem}
\setcitestyle{square,numbers,sort&compress,comma}
\newcommand{\str}[1]{} 

\newcommand{\tr}[1]{\textcolor{black}{#1}} 
\newcommand{\bayu}[1]{\textcolor{black}{#1}} 

\date{January 2023}

\newcommand{\OHstar}{$\mathrm{OH^{*}}$}

\begin{document}
\begin{frontmatter}
\title{Thermoacoustic Stabilization of a Sequential Combustor with \tr{Ultra-}low-power Nanosecond Repetitively Pulsed Discharges}
\author{Bayu Dharmaputra\corref{cor1}}
\ead{bayud@ethz.ch}
\author{Sergey Shcherbanev\corref{}}
\author{Bruno Schuermans\corref{}}

\author{Nicolas Noiray\corref{cor1}}
\ead{noirayn@ethz.ch}
\cortext[cor1]{Corresponding authors}

\address{CAPS Laboratory, Department of Mechanical and Process Engineering, ETH Z\unexpanded{\"u}rich, 8092, Z\unexpanded{\"u}rich, Switzerland}
\begin{abstract}

 This study demonstrates the stabilization of a sequential combustor with Nanosecond Repetitively Pulsed Discharges (NRPD). A constant pressure sequential combustor offers \str{many} \tr{key} advantages compared to a conventional combustor, \str{including} \tr{ in particular, a} higher fuel flexibility and a wider operational range. However, thermoacoustic instabilities remain a barrier to \tr{further} widen\str{ing} the operational range of the\tr{se} combustor\tr{s}. \str{In the past decades, both active and passive control strategies for gas turbine combustors have been studied.}Passive control strategies \tr{to suppress these instabilities, such as Helmholtz dampers,} have been \str{more widely used in commercial} \tr{used in some industrial} systems \tr{thanks} to their simplicity in terms of implementation. Active control strategies \tr{are however not found in practical combustors}, mainly due to the lack of robust actuators \tr{able to operate in harsh conditions} with sufficient control authority. In this study, we demonstrate \tr{that thermoacoustic instabilities can be suppressed by using a non-equilibrium plasma produced with} NRPD \tr{in} a lab-scale atmospheric sequential combustor operated at 73.4 kW of thermal power. We employ continuous \str{forcing of} NRPD \tr{forcing} to \tr{influence} the combustion \tr{process in} the sequential combustor. The two governing parameters are the \str{plasma}\tr{pulse} repetition frequency (PRF) and the plasma generator voltage. We examine the effect of both parameters on the \str{pressure pulsation}\tr{acoustic amplitude}, the NO emissions, and the flame centre of mass. We observe that \tr{for some operating conditions,} with plasma power \tr{of 1.1 W, which is about}\str{as low as} $\str{1.4}\tr{1.5} \times 10^{-3}\str{\%}$ \tr{percent} of the thermal power of the flames, the combustor can be thermoacoustically stabilized. By increasing the power of the plasma, \tr{the acoustic amplitude} can be \tr{further reduced, at a small cost of} a very low increase of NO emission. However, \tr{an additional} increase in plasma power to 81~W, which is $\str{1.4}\tr{1.1} \times 10^{-1}\str{\%}$ \tr{percent} of the flame thermal power, increase the NO emission significantly without significant improvement on the \tr{acoustic amplitude} reduction. \tr{Furthermore, for} some combination\tr{s} of the plasma parameters, another \tr{thermoacoustic} mode of the combustor at a different frequency can be\tr{come unstable}\tr{. This finding motivates further research on the} optimization of the \tr{plasma} parameters \tr{as a function of the thermoacoustic properties of the combustor where it is applied}. This study is a pioneering effort in controlling the thermoacoustic stability of turbulent flames with \bayu{plasma discharges} at such low power compared to the thermal power of the \tr{sequential combustor.}

\end{abstract}

\end{frontmatter}
\newpageafter{abstract}

\section*{Novelty and significance}

In this study, we demonstrate that Nanosecond Repetitively Pulsed Discharges (NRPD) can be used as an actuator to stabilize a thermoacoustically unstable sequential combustor. We suppress instabilities with a mean plasma power that is 5 orders of magnitude lower than the flames thermal power, i.e. more than 2 orders of magnitude lower than the current state of the art. This achievement therefore opens the door for commercial application of such a technology for continuous operation. Furthermore, we examine the impact of varying plasma parameters, specifically the plasma repetition frequency and generator voltage, on flame topology changes, as well as investigate the resulting NO emissions.

\section*{Authors contributions}

B.D. led the experimental investigations and designed the analyses. B.D. and S.S. performed the experiments and analysed the data. N.N. conceived the research idea. B.D., S.S., B.S. and N.N. discussed the results. B.D. drafted the manuscript with the support of N.N. The final version of the manuscript has been edited and approved by B.D., S.S., B.S. and N.N.

\section{Introduction}

Gas turbines have played a significant role in global energy production. However, with the increasing proportion of renewable energy production and tightening emission restrictions, gas turbines now face \str{additional} \tr{new} challenges, \str{including} \tr{and in particular} the need for fuel flexibility, fast ramp-up times, and a wide operational range \cite{Ciani2019}. One major technological breakthrough \tr{of the recent past in} \str{ for} gas turbine \str{applications} \tr{technologies} is the constant pressure sequential combustor (CPSC) \cite{Pennell2017, Bothien2019}.\str{, which} \tr{It} significantly improves \tr{the} operational range \tr{with very low pollutant emissions,} and \tr{the }fuel flexibility\str{\cite{Ciani2019}. Fuel flexibility includes}\tr{, which is} the ability to \tr{be supplied with} blend\tr{s of} hydrogen and natural gas, as well as \str{the use of} non-conventional fuels\str{, such as those} derived from waste processes or biomass gasification \cite{Ciani2019}. \tr{Notably, the combustion of pure H$_2$ in an academic CPSC configuration has been investigated recently in \cite{SOLANAPEREZ2022}.} In \str{the} CPSC configuration\tr{s}, the hot products from the first-stage combustor are diluted with \str{additional fresh} air \tr{bypassing the first-stage before the sequential fuel is injected.}\str{, which} \tr{This combustor architecture} reduces the \str{flow} temperature \tr{of the vitiated air flow at the inlet of the sequential stage in order} to prevent \str{instantaneous}\tr{too fast auto}ignition of the \str{secondary}\tr{sequential} fuel \str{near the injection region}\tr{in poorly mixed conditions. Ensuring in this way a sufficient mixing time of the globally lean mixture of gas and vitiated air, the exothermal reactions of the autoignition process in the sequential stage  occur in well-mixed conditions, and the NO$_\mathrm{x}$ emissions can thus be drastically reduced.} \str{Consequently, combustion in the sequential flame relies mostly on the autoignition mechanism \cite{SOLANAPEREZ2022,Schulz2019}.}


Much like a traditional combustor, CPSC \str{is}\tr{combustors are} also prone to thermoacoustic instabilities \cite{Schulz2019, Bonciolini_ASME}. Thermoacoustic instabilities are challenging problems in gas turbines for power and propulsion applications~\cite{Lieuwen2005}. These instabilities can \str{manifest}\tr{lead to high amplitude acoustic} pressure \str{fluctuations}\tr{oscillations which can induce vibrations causing structural damages and possibly} \str{reaching a large fraction of the mean pressure which could lead to} flame flashback\str{, and in the extreme cases to structural damage}~\cite{Poinsot2017}. Hence, \tr{developing technologies for} controlling the instabilities is \str{an} essential \str{task} for the safety and operability of gas turbines. In the past decades, both active and \tr{passive} control strategies have been \str{studied}\tr{investigated in academic and industrial laboratories and implemented in real engines. Nonetheless, gas turbine manufacturers  usually opt for passive control strategies which are more cost-effective so far}. 


\tr{Indeed, }passive damping strategies have been widely studied and applied in real combustors\tr{. For example, the nonlinear behavior of Helmholtz resonators mounted on combustion chamber walls has been  investigated in a recent study in order to  draw design guidelines for avoiding failures of their damping effectiveness} \cite{Miniero2023}. \str{For example, a Helmholtz-type damper was implemented in the sequential combustor of the Ansaldo GT26}\tr{Furthermore, dampers based of interconnected cavities with broadband acoustic absorption capabilities were successfully implemented in large modern gas turbines} \cite{bothien_2014,Zahirovic2017}. However, \str{such a passive strategy} \tr{the design of these passive dampers is still challenging for the following two reasons: First, it}  requires \tr{costly engine testing for obtaining a} relatively precise \tr{prior} knowledge of the \str{system's eigenmodes}\tr{difficult-to-predict thermoacoustic instabilities,}\str{in advance} in order to \tr{tune their geometry for} effective\str{ly} reduc\str{e}\tr{tion of the acoustic}\str{pulsation} amplitude. \tr{Second, for a given volume constraint for their implementation, there is always a trade-off to find between their broadbandness for addressing multiple instability frequencies and their effectiveness at a given frequency.}

In contrast, active control strategies\str{, with proper parameter tuning,} can adapt to the operating conditions of the system \tr{but, so far, their implementation in real engines has been hindered by the harsh thermodynamic and thermochemical conditions and  by the lack of cost-effective and mechanically-robust actuation solutions}. The \tr{big} challenge of implementing active control strategies\str{, however,} is \tr{thus} finding suitable actuators \tr{\cite{Docquier2002107,Dowling2005}}. For example, \str{\cite{Moeck2010} demonstrated} the use of \str{pilot fuel flow modulation and} loudspeaker\tr{s} \str{forcing} to stabilize an unstable combustor \tr{by tailoring its acoustic boundary conditions was successfully achieved in an academic configuration operated at atmospheric pressure} \str{. However, as the author noted, using loudspeakers has a scalability limitation at high pressures. Moreover, pilot fuel modulation can lead to an increase in NO emissions.} \tr{\cite{Bothien2008678} but could not be applied in a real engine.} \tr{Another active control strategy, based on the fast modulation of the fuel mass flow  has been successfully developed  about thirty years ago for liquid spray \cite{hermann_1996} and natural gas \cite{Seume1998}. The latter technology has even been validated in heavy duty gas turbines and was commercialised. Nonetheless, effective modulation the pilot gas massflow cannot be achieved beyond 500 Hz, which prevents from addressing the problem of high-frequency instabilities with the corresponding valves. Furthermore, such modulation of the fuel massflow could negatively impact the pollutant emissions.}

 \str{Therefore, finding an} \tr{In this context, the search of alternative} actuator\tr{s for gas turbine combustors,} with high control authority, low power consumption, and minimal additional emissions\str{, remains a challenge} \tr{is highly relevant for increasing fuel and operational flexibility of future gas turbines. In this work, we focus on  ultra-low-power plasma actuation, which has never been implemented in industrial systems so far, and we show that it is a very promising strategy for suppressing thermoacoustic instabilities in sequential combustors without  increasing NO$_\mathrm{x}$ emissions}.

Non-equilibrium plasma discharges \str{have been shown to have several benefits when it comes to}\tr{can be used to enhance}\str{enhancing and stabilizing} combustion \str{processes} \tr{reactions through thermochemical effects}~\cite{starikovskaia2014plasma,ju2015plasma,starikovskiy2015physics,adamovich2014challenges, LACOSTE2022}. \str{First, they can be used to enhance the ignition of fuel mixtures by creating free radicals and other active species. A}\tr{In these plasmas, the} substantial difference between electron and gas temperatures ($T_{e} \gg T_{gas}$) results in the efficient formation of active species and radicals through direct electron impact, which can help to \tr{ignite combustible mixtures, and to extend}\str{initiate and promote combustion reactions. This contributes to a more complete combustion process and an extension of the} lean flammability limit\tr{s}~\cite{pilla2006stabilization,bak2012plasma,patel2023low}. \str{Second, non-equilibrium plasma discharges can also help to stabilize the combustion process. This is because the plasma can provide a steady stream of active species to the system, which can help to maintain a stable flame. This can be particularly useful in lean or unstable combustion environments, where the flame can easily be blown off or become unstable.}

Owing to its high influence to the kinetics of the reactive mixture, several studies have been \str{done to investigate} \tr{performed to characterize} the effect of \tr{nanosecond repetitively pulsed discharges (}NRPD\tr{)} on the \str{flame} heat release rate \tr{oscillations of acoustically-forced  flames. For instance,} \str{response.}Lacoste et al. \cite{Lacoste2013} \str{investigates the effect of plasma actuation to the heat release rate fluctuation with respect to the acoustic perturbation} \tr{have studied this effect} in a single stage swirled stabilized combustor. It was shown that NRPD affects significantly the gain and phase \tr{of the }flame transfer function (FTF) and thereby, might influence the \tr{thermoacoustic} stability of the combustor. In \cite{Lacoste2017}\tr{, it was shown with a laminar flame }that \tr{strong} heat release rate \tr{modulation} \str{fluctuation of the laminar flame responds strongly under the influence of} \tr{can be induced by}  periodic \tr{series of constant voltage} NRPD\str{ with square-wave input}. The forcing mechanism was mainly attributed to the increase of local burning velocity close to the plasma region. In a similar set-up to the one in \cite{Lacoste2013}, Moeck et al.~\cite{Moeck2013} \str{has} \tr{have} successfully demonstrated the applicability of nanosecond plasma discharges to stabilize a linearly unstable combustor with active feedback control. By using the extended kalman filter (EKF), the instantaneous phase of the acoustic pulsation was estimated and then fed to the gate signal for the actuation of the plasma generator. The plasma power required to stabilize the combustor was at around $1\str{\%}$ \tr{percent} of the flame thermal power. \bayu{Furthermore, Kim et al \cite{Kim2021} have demonstrated the capability of NRPD to stabilize a combustor at realistic low power condition of aero-engine combustors.}


Nanosecond plasma discharges in pin-to-pin configuration have shown \tr{high} potential \str{as a tool} for \tr{second-stage} flame stabilization in constant-pressure sequential combustors, as  \str{they offer several advantages over traditional flame stabilization methods}\tr{ the hot reactive mixture in the sequential burner already undergoes radicals-producing chemical reactions that precede autoignition}. Xiong et al. \cite{Xiong2019} demonstrated that NRPD could shorten significantly the auto-ignition time of \str{the} \tr{a CH$_4$} sequential flame with low electric power and \str{acceptable} NO emissions. In a \tr{more compact laboratory-scale sequential combustor}, Shcherbanev et al. \cite{Shcherbanev2022} demonstrated the effectiveness of plasma discharges in igniting \tr{very} lean mixture\tr{s} of hydrogen and natural gas\str{blending}, and therefore could help keeping the sequential flame alive. One \tr{significant} advantage of the NRPD \tr{actuation} is \tr{that} the electrode system \tr{implementation} does not require \tr{a drastic} modification of the combustor geometry \tr{and could also  be considered for} retrofitting existing systems. 

Another advantage of NRPD is their fast response time\str{and robustness. They do not require any}\tr{, which enables high frequency actuation without} moving \tr{parts}\str{components, which} \tr{that are usually causing reliability and durability issues in mechanical actuators. 
Indeed, non-equilibrium}\str{increases their reliability and durability. The}  plasma actuation can quickly \str{respond to fluctuations in the} \tr{influence}  flame\tr{s,}\str{, resulting in} improv\str{ed flame}\tr{ing their}  stability and extending \tr{their} lean flammability limit~\cite{Lacoste2013b,DiSabatino2020}. 

Finally,  \bayu{the mean NRPD power is known to be small compared to the thermal power of the flame} ~\cite{Kim2015,Lacoste2013b,Lacoste2013}. \bayu{It should be noted that the mean NRPD power reported in literature pertains only to the electrical energy that gets transferred to the system, without accounting for the electrical energy necessary for the high voltage generator. However, if we consider an optimistic scenario where an exceptionally efficient high voltage generator is feasible, including the achievement of a perfect impedance matching at the electrodes, the generator's power requirements would match the energy deposited into the system.} \tr{In the case of a sporadic use of the NPRD, for flame ignition assistance or during transient operation, the electric power requirements of a NRPD system is not a major driver in the development of plasma assisted combustion technologies. Although the mean NRPD electric power reported in previous works is of the order of 1 percent of the thermal power of the flame \cite{Moeck2013, Alkhalifa2022}, it is worth mentioning that the cost-benefit 
analysis of a NRPD system with an electric power of 1 percent of the combustor thermal power would not be straightforward for heavy duty gas turbines, for the following two reasons. First, 1~percent is still a large penalty for gas turbine manufacturers which struggle for gaining any 0.1 percent of engine thermal efficiency during stationary operation (over the last  decade, this efficiency increased toward 65 percent for combined cycle power plants by only a couple of percents). Second, for an H-class gas turbine exhibiting a combustor with 1 GW of thermal power, it means that, \bayu{by assuming linear scaling}, a system of 10 MW electric power would have to be developed, just for the NRPD actuation, which is technically rather challenging. In the present work, we demonstrate that for a sequential combustor, operated at atmospheric pressure, successful actuation suppressing thermoacoustic instabilities can be achieved with a \bayu{mean} plasma power that is about 3 orders of magnitude lower than 1 percent of the thermal power, which would be much more realistic for implementation in practice (for 1 GW of thermal power, one would need about 10 kW of \bayu{mean} plasma power). Therefore, a}s research in this area continues, it is \str{likely}\tr{foreseen} that \str{this}\tr{such NRPD} technology \str{will become increasingly prevalent in the design and operation of}\tr{may be implemented in future} gas turbines \tr{burning green H$_2$ in sequential combustors for compensating the intermittency of renewable sources.}\str{, helping to reduce emissions and improve sustainability in the energy sector.} 

\str{However, no study has attempted to utilize the plasma discharges for stabilizing a thermoacoustically unstable sequential combustor. Since NRPD could influence the flame position and helps anchoring the flame in a sequential combustor \cite{Shcherbanev2022}, it is natural to hypothesize that the thermoacoustic stability could be influenced as well.  }


This study \tr{thus} aims at introducing the \tr{ultra-}low-power nanosecond plasma rapid discharges (NRPD) to thermoacoustically stabilize a sequential combustor \tr{which had never been attempted so far}. Additionally, parametric studies on the plasma repetition frequency (PRF) and the generator voltage are \str{done}\tr{performed} to investigate the effectiveness of the actuator \tr{in suppressing the instability }and the \tr{associated} NO emissions.


\section{Experimental Setup}

\begin{figure}[t!]
	\centering
	\includegraphics[width=1.0\linewidth]{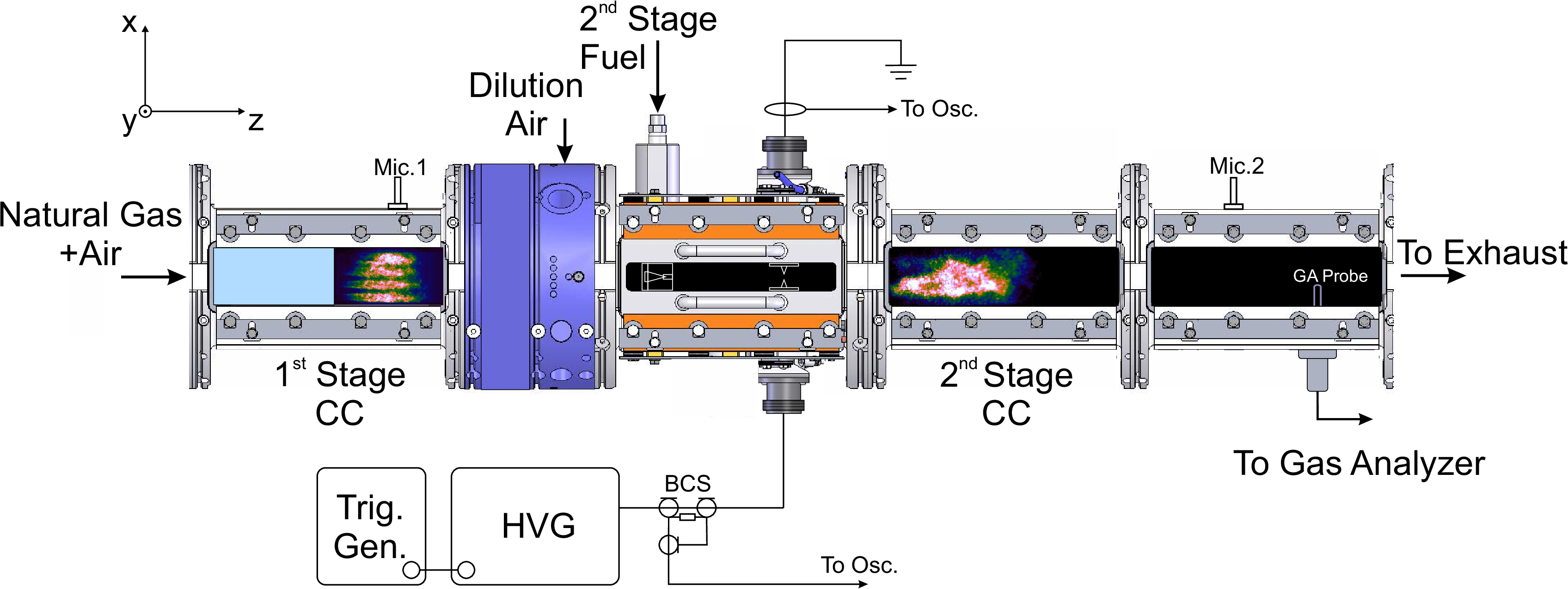}   
	\caption{Lab-scale sequential combustor test-rig. CC- Combustion chamber, \bayu{HVG- High Voltage Generator, BCS- Back Current Shunt.}}
    \label{fig:testrig}
\end{figure}

The lab-scale sequential combustor \str{setup} is depicted in figure \ref{fig:testrig}. The setup consists of a plenum, a 4 $\times$ 4 \tr{array of jet flames anchored on a so-called} matrix burner, a combustion chamber with 62 $\times$ 62 $\mathrm{mm^2}$ cross section, a dilution \tr{air} section, a \tr{sequential burner featuring a }mixing channel with 25 $\times$ 38 $\mathrm{mm^2}$ cross section\str{, a}\tr{in which}  secondary fuel \str{injection}\tr{is injected}, a sequential or second-stage combustion chamber \str{, and}\tr{equipped with a motor-driven} adjustable \tr{outlet} orifice\tr{. This variable outlet geometry enables an online tuning of the acoustic reflection coefficient, and thus  an independent control of the thermoacoustic instabilities, which is key for validating the NRPD-based control}. The first stage combustor \str{uses}\tr{is fed with} a mixture of natural gas and air, with the air preheated to 230 C and \str{introduced}\tr{supplied} from the plenum, while natural gas is added \str{through}\tr{in} the matrix burner, \str{creating a}\tr{which corresponds to a } technically premixed \str{mixture}\tr{first stage}. The thermal power of the first stage combustor is 35~kW with an equivalence ratio of 0.7. A piezo sensor is placed on a flush mounted plate to monitor the \tr{acoustic }pressure \str{pulsation} inside the first stage \str{combustor} and denoted as Mic.~1 in the figure. \tr{A massflow of }18~g/s of dilution air at 25~C is introduced from the dilution air port and mixes with the hot gases from the first \str{combustor}\tr{stage}. A mixture of 0.07~g/s \tr{of} hydrogen and 0.6~g/s  \tr{of} natural gas is injected \tr{into} the sequential injector. The sequential injector features an X\tr{-}\str{lobe}shaped vortex generator to \str{introduce rotational movement to the vitiated flow which} enhance\str{s} the mixing process. The total thermal power of the two flames is 73.4~kW. A pin-to-pin electrode configuration, with an inter-electrode distance of 5~mm, is located 10.3~cm \tr{downstream }from the sequential fuel injector, and a gas analyzer (ABB-EL3040) probe is placed at 45~cm from the  outlet of the second-stage  \str{combustor}\tr{burner} to monitor the NO emissions. Another piezo sensor is placed downstream of the sequential flame to monitor the \tr{acoustic }pressure \str{pulsation of}\tr{in} the second combustion chamber.  \tr{As indicated above, t}\str{T}he outlet  \tr{of this chamber} has an adjustable orifice to  \str{change}\tr{control} the \str{nominal}thermoacoustic stability\str{of the system}. 

The \OHstar{} chemiluminescence is used to characterized the sequential flame, with the camera capturing a portion of the mixing channel downstream of the electrodes.\str{However, due to t}\tr{T}he intense \tr{light} emission \str{s}from the plasma discharges \str{,}  \tr{is masked by}an optical  \str{screen is  required to mask them} \tr{obstacle}. The recording setup comprises of a LaVision Star X high-speed CMOS Camera and a LaVision HS-IRO high-speed intensifier, which are equipped with a 45 mm CERCO UV lens (F$/$1.8 Cerco) and an Edmund Optics optical bandpass filter (centered at 310~nm, FWHM~10nm).

To measure the energy deposition of the plasma, a current probe and a back current shunt are \tr{placed} on the anode and cathode to \tr{acquire} \str{measure}current and voltage. The plasma generator (FID) initiates the high voltage pulses with \tr{a} 2-3~ns rise time and a pulsed width of 10~ns. \tr{A }\str{Tektronix (Tektronix MDO3104)}mixed signal digital oscilloscope \tr{(Tektronix MDO3104)} was used to record the current and the voltage signals at 1~GHz bandwidth and 5~GHz sampling rate. A Pearson fast current monitor (model 6585, 0.5 V/A, 50.) \str{were}\tr{was} used to measure the  current. During testing, four generator voltages (11, 11.7, 12.5, and 14~kV) at three pulse repetition frequencies (10, 20, 40~kHz) were used, with an additional voltage value of 13.2~kV at 40~kHz. All \bayu{experiments} were performed using only the  negative polarity \bayu{of applied pulses}.

\begin{figure}[t!]
    \centering
    \begin{psfrags}

    \psfrag{NOemission}[][]{\hspace{-0.6cm}\scriptsize{NO~($\mathrm{mg/mm^3}$)}}
    \psfrag{COMmm}[][]{\scriptsize{COM (mm)}}
    
    \psfrag{WithPlasma}[][]{\scriptsize{with plasma}}
    \psfrag{WithoutPlasmaxxx}[][]{\hspace{-0.4cm}\scriptsize{without plasma}}    
    \psfrag{0}[][]{\scriptsize{0}}    
    \psfrag{0b}[][]{\color{blue}\scriptsize{0}}   
    \psfrag{-2.5}[][]{\color{blue}\scriptsize{-2.5~~}}
    \psfrag{-5}[][]{\color{blue}\scriptsize{-5~~}}
    \psfrag{2.5}[][]{\color{blue}\scriptsize{2.5~~}}
    
    \psfrag{0r}[][]{\color{red}\scriptsize{0}}
    \psfrag{-4000}[b][]{\color{red}\scriptsize{\hspace{0.01cm}-4}}
    \psfrag{-2000}[][]{\color{red}\scriptsize{\hspace{0.01cm}-2}}
    \psfrag{2000}[][]{\color{red}\scriptsize{\hspace{0.01cm}2}}

    \psfrag{30}[][]{\scriptsize{30}}
    \psfrag{40}[][]{\scriptsize{40}}
    \psfrag{50}[][]{\scriptsize{50}~}
    \psfrag{-50}[][]{\scriptsize{-50}~~}
    \psfrag{-100}[][]{\scriptsize{-100}~~~}
    
    \psfrag{60}[][]{\scriptsize{60}}
    \psfrag{80}[][]{\scriptsize{80}}    
    \psfrag{100}[][]{\scriptsize{100}}
    \psfrag{150}[][]{\scriptsize{150}}
    \psfrag{200}[][]{\scriptsize{200}}    
    \psfrag{250}[][]{\scriptsize{250}}   
    \psfrag{timens}[t][]{\scriptsize{$t$~(ns)}}  
    \psfrag{VoltagekV}[t][]{\color{red} \scriptsize{$V$~(kV)}}  
    \psfrag{CurrentmA}[t][]{\color{black} \scriptsize{$I$~(A)}}  
    \psfrag{EnergyDeposmJ}[b][]{\color{blue} \scriptsize{$E_{D}$~(mJ)}}  

    \psfrag{Voltage}[][]{\color{black} \scriptsize{~~~~Voltage}}  
    \psfrag{Current}[][]{\color{black} \scriptsize{~~~~Current}}  
    \psfrag{EnergyDepositionxxxxxxx}[][]{\color{black} \scriptsize{Energy deposition}}
    
    \includegraphics[trim=0cm 0cm 0cm 0cm,clip,width=0.55\textwidth]
    {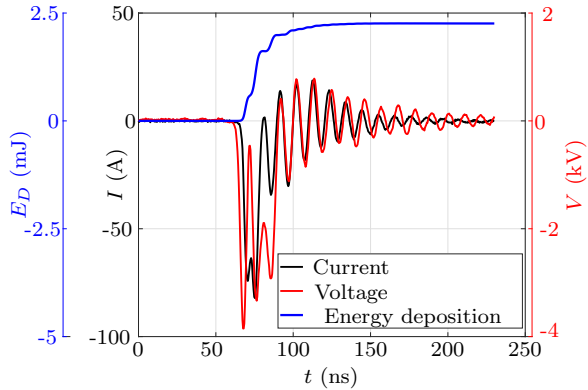}
    \end{psfrags}
    \caption{Waveforms of current (black), voltage (red), and energy deposition (blue) . The energy deposition is obtained by taking the integral of the product of the current and voltage signal. \bayu{The measured voltage corresponds to the voltage across the electrodes' gap and hence it is not necessarily equal to the voltage of the generator.}}
    \label{fig:EnergyDeposition}
\end{figure}

\begin{figure}[t!]
    \centering
    \begin{psfrags}

    \psfrag{NOemission}[][]{\hspace{-0.6cm}\scriptsize{NO~($\mathrm{mg/mm^3}$)}}
    \psfrag{COMmm}[][]{\scriptsize{COM (mm)}}
    
    \psfrag{WithPlasma}[][]{\scriptsize{with plasma}}
    \psfrag{WithoutPlasmaxxx}[][]{\hspace{-0.4cm}\scriptsize{without plasma}}    
    \psfrag{0}[][]{\scriptsize{0}}    
    \psfrag{0b}[][]{\color{blue}\scriptsize{0}}   
    \psfrag{-2.5}[][]{\color{blue}\scriptsize{-2.5~~}}
    \psfrag{-5}[][]{\color{blue}\scriptsize{-5~~}}
    \psfrag{2.5}[][]{\color{blue}\scriptsize{2.5~~}}
    
    \psfrag{0r}[][]{\color{red}\scriptsize{0}}
    \psfrag{-4000}[b][]{\color{red}\scriptsize{\hspace{0.01cm}-4}}
    \psfrag{-2000}[][]{\color{red}\scriptsize{\hspace{0.01cm}-2}}
    \psfrag{2000}[][]{\color{red}\scriptsize{\hspace{0.01cm}2}}

    \psfrag{11}[][]{\scriptsize{11}}
    \psfrag{12}[][]{\scriptsize{12}}
    \psfrag{13}[][]{\scriptsize{13}}
    \psfrag{14}[][]{\scriptsize{14}}

    \psfrag{1e-3}[][]{\scriptsize{$10^{-3}$}}
    \psfrag{1e-1}[][]{\scriptsize{$10^{-1}$}} 
    \psfrag{1e-2}[][]{\scriptsize{$10^{-2}$}}    
    \psfrag{1e1}[][]{\scriptsize{$10^1$}}
    \psfrag{1e3}[][]{\scriptsize{$10^3$}}
    \psfrag{1e2}[][]{\scriptsize{$10^2$}}
    \psfrag{1e0}[][]{\scriptsize{1}}
    
    \psfrag{1.4e-4}[][]{\scriptsize{$~~~~~\str{1.4}\tr{1.5}\times 10^{-4}$}}
    \psfrag{1.4e-3}[][]{\scriptsize{$~~~~~\str{1.4}\tr{1.5}\times 10^{-3}$}}
    \psfrag{1.4e-2}[][]{\scriptsize{$~~~~~\str{1.4}\tr{1.5}\times 10^{-2}$}}
    \psfrag{1.4e-1}[][]{\scriptsize{$~~~~~\str{1.4}\tr{1.5}\times 10^{-1}$}}
    \psfrag{1.4e0}[][]{\scriptsize{\str{1.4}\tr{1.5}~~~~}}
    \psfrag{RatioPercent}[t][]{\scriptsize{$\tr{\eta_p}~(\%)$~~~~~~~~~~~~}}  
    
    \psfrag{a}[][]{\scriptsize{a)}}
    \psfrag{b}[][]{\scriptsize{b)}}
    
    \psfrag{10kHzxxxx}[][]{\hspace{-0.75cm}\scriptsize{10 kHz}} 
    \psfrag{20kHz}[][]{\scriptsize{20 kHz}}    
    \psfrag{40kHz}[][]{\scriptsize{40 kHz}}
    
    \psfrag{timens}[t][]{\scriptsize{$t$~(ns)}}  
    \psfrag{VoltagekV}[t][]{\color{black} \scriptsize{$V$~(kV)}}  
    \psfrag{EnmJ}[][]{\color{black} \scriptsize{$E_D$~(mJ)}}  
    
        \psfrag{powerwatt}[][]{\color{black} \scriptsize{Power (W)}} 
    
    \psfrag{CurrentmA}[][]{\color{black} \scriptsize{$I$~(A)}}  
    \psfrag{EnergyDeposmJ}[][]{\color{blue} \scriptsize{$E_{D}$~(mJ)}}

    \psfrag{Voltage}[][]{\color{black} \scriptsize{~~~~Voltage}}  
    \psfrag{Current}[][]{\color{black} \scriptsize{~~~~Current}}  
    \psfrag{EnergyDepositionxxxxxxx}[][]{\color{black} \scriptsize{Energy deposition}}
    
    \includegraphics[trim=0cm 0cm 0cm 0cm,clip,width=1\textwidth]
    {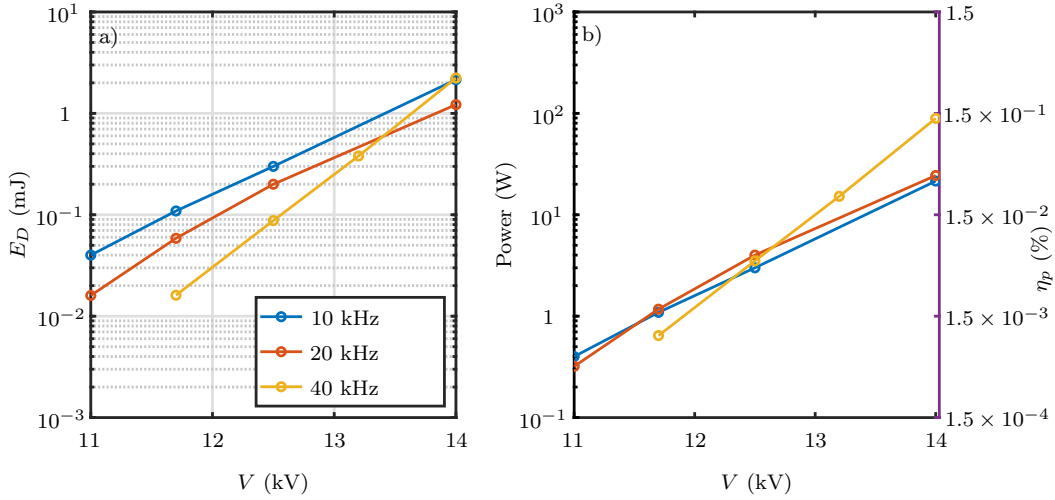}
    \end{psfrags}
    \caption{Energy deposition measurement a) and the plasma power b) at three different PRFs and four generator voltages. The \bayu{mean} plasma power is obtained by taking the product of the PRF and the energy deposition. The ratio between the \bayu{mean} plasma power and the thermal power of the combustor is indicated on the right y-axis in b). There is no plasma discharge observed at PRF =40~kHz and $V$ = 11~kV.}
    \label{fig:EnergyDepositionvsfreq}
\end{figure}

The energy deposition measurement \tr{of a single pulse for continuous NRPD }at 14 kV and 10 kHz is shown in figure \ref{fig:EnergyDeposition}. Note that due to the mismatch between the generator impedance and the plasma impedance, oscillations in voltage and current are observed. The deposited energy is computed by taking the time integral of the product \str{between}\tr{of} voltage and current. At this condition, mean energy deposition is around 2~mJ \tr{per pulse}. The type of plasma at this energy is classified as spark plasma~\cite{Minesi2022}. The resulting energy deposition at all conditions are shown in figure \ref{fig:EnergyDepositionvsfreq}a. \str{Note}\tr{It is interesting to note that the y-axis has a} \str{the} logarithmic scale\str{of the y-axis}, \tr{and that} the energy deposition increases exponentially with respect to the generator voltage. Furthermore, except at 14 kV, the energy deposition \tr{per pulse} at higher frequency and at the same voltage decreases. This\str{could} \tr{is} probably \str{be attributed}due to the interference between the incoming and reflected pulses inside the high voltage cable. Furthermore, there was no plasma observed at 11 kV and 40 kHz and consequently the data can not be shown in the plot.

By multiplying the energy deposition and the plasma repetition frequency (PRF), the \bayu{mean} plasma power is obtained and depicted in figure \ref{fig:EnergyDeposition}b. The highest power is at \str{80}\tr{81} W which amounts to \str{1.4}$\tr{1.1}\times 10^{-1}\str{\%}$ \tr{percent} of the thermal power of the flames. The ratio between the plasma to thermal power is indicated on the right axis of the figure \ref{fig:EnergyDepositionvsfreq}b) which we denote as $\tr{\eta_p}$.

\section{Results}

\begin{figure}[t!]
    \centering
    
    \psfrag{aaa}[][]{\scriptsize a~~}
    \psfrag{bbb}[][]{\scriptsize b~~}
    \psfrag{ccc}[][]{\scriptsize c~~}
    \psfrag{ddd}[][]{\scriptsize d~~}
    \psfrag{eee}[][]{\scriptsize e~~}
    \psfrag{fff}[][]{\scriptsize f~~} 
    
    \psfrag{a}[][]{\scriptsize a)}
    \psfrag{b}[][]{\scriptsize b)}
    \psfrag{c}[][]{\scriptsize c)}
    \psfrag{d}[][]{\scriptsize d)}
    \psfrag{e}[][]{\scriptsize e)}
    \psfrag{f}[][]{\scriptsize f)}    
    \psfrag{g}[][]{\scriptsize g)}
    \psfrag{h}[][]{\scriptsize h)}
    \psfrag{0.5}[][]{}
    \psfrag{timeSec}[][]{\scriptsize $t$ (s)}
     \psfrag{FreqHz}[][]{\scriptsize $f$ (Hz)}
    \psfrag{Pressure}[][]{\scriptsize p (mbar)}
    \psfrag{PlasmaONNNNN}[][]{\scriptsize W/o plasma}
    \psfrag{PlasmaOFFFFFF}[][]{\scriptsize W/ plasma}
    \psfrag{mbar}[][]{\scriptsize mbar}
    \psfrag{powerdba}[c][t][0.8]{\scriptsize $S_{pp} (dBa)$}
    \psfrag{scaledpdf}[l][t][1][270]{\scriptsize $\hat{P_{p}}$}
    
    \psfrag{pi2}[][]{\scriptsize $\frac{\pi}{2}$}
    \psfrag{pi3}[][]{\scriptsize $\pi$}
    \psfrag{pi4}[][]{\scriptsize $\frac{3\pi}{2}$}
    \psfrag{pi4}[][]{\scriptsize $\frac{3\pi}{2}$}
    \psfrag{min}[][]{ \scriptsize{min}}
    \psfrag{max}[][]{\scriptsize{max}}

    \psfrag{0}[][]{\scriptsize{0}}
    \psfrag{250}[][]{\scriptsize{250}}
    \psfrag{500}[][]{\scriptsize{500}}

    \psfrag{750}[][]{\scriptsize{750}}
    
    \psfrag{20}[][]{\scriptsize{20}}
    \psfrag{40}[][]{\scriptsize{40}}
    \psfrag{60}[][]{\scriptsize{60}}

    \psfrag{110}[][]{\scriptsize{110}~~}
    \psfrag{140}[][]{\scriptsize{140}~~}
    \psfrag{170}[][]{\scriptsize{170}~~}

    \psfrag{100}[][]{\scriptsize{100}}
    \psfrag{-50}[][]{\scriptsize{-50}}
    \psfrag{50}[][]{\scriptsize{50}}
    \psfrag{-200}[][]{\scriptsize{-200}}
    \psfrag{-350}[][]{\scriptsize{-350}}
    \psfrag{-500}[][]{\scriptsize{-500}}
    \psfrag{-100}[][]{\scriptsize{-100}}
    \psfrag{-150}[][]{\scriptsize{-150}}    
    \begin{tikzpicture}
       \node[anchor=south west,inner sep=0] at (0,0) {    \includegraphics[trim=0cm 0cm 0cm 0cm,clip,width=1\textwidth]{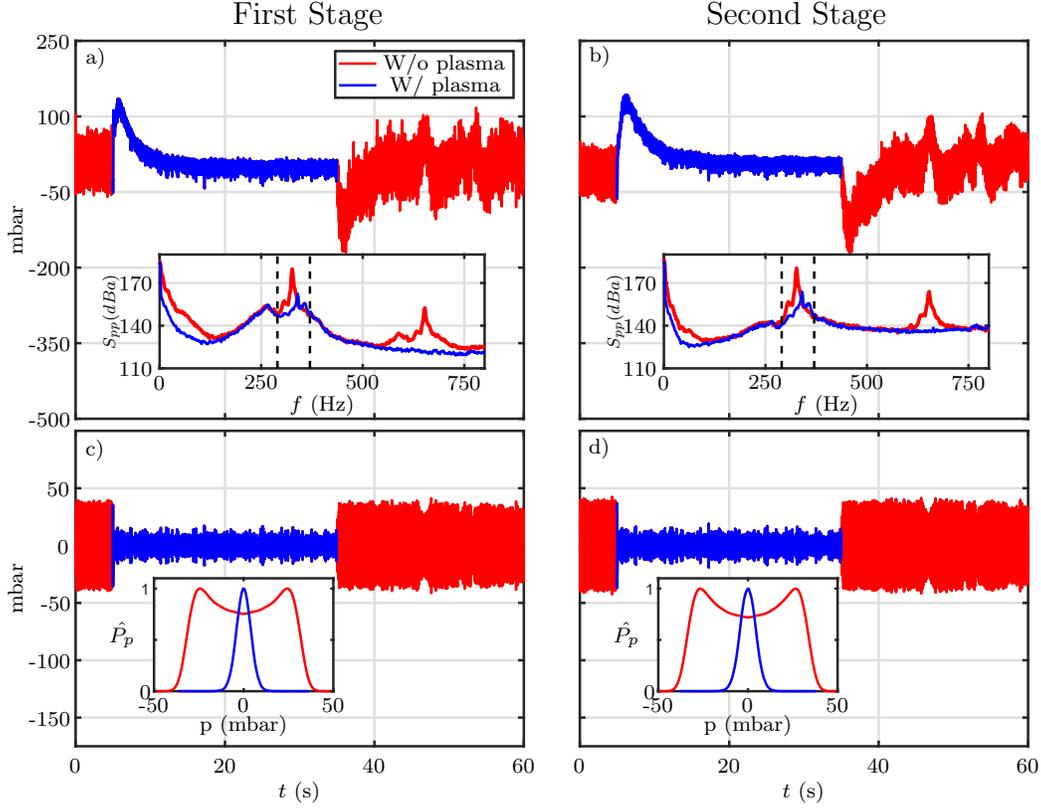}};
       \node[align=center] at (4,10.1) {\bayu{First Stage}\\};
       \node[align=center] at (10.5,10.1) {\bayu{Second Stage}\\};
   \end{tikzpicture}
    
    \caption{The time trace of the acoustic pressure signal with and without plasma discharges and their corresponding \str{frequency}\tr{power} spectra\tr{l density} inside the first a) and second b) stage \str{combustors}\tr{combustion chambers}. The PRF is at 10 kHz and the generator voltage is at 11.7 kV, the \bayu{mean} plasma power is 1.1 W\tr{, which is about 1.5 $\times~10^{-3}$ percent of the thermal power of the flame}. The bandpass filtered acoustic pressure signal and its associated normalized probability density function inside the first c) and second d) stage \str{combustors}\tr{combustion chambers}. The low-pass and high-pass frequency are indicated by the dashed lines in a) and b).}
    \label{fig:micdata}
\end{figure}

Figure \ref{fig:micdata} shows the effects of plasma discharges on the stability of the sequential combustor at a PRF of 10~kHz and generator voltage of 11.7~kV. The plasma was turned on at $t=5$~s and turned off at $t=35$~s. In Figure \ref{fig:micdata}a and \ref{fig:micdata}b, the \tr{acoustic} pressure signals from the first and second stage \str{combustors}\tr{combustion chambers} are presented, along with their corresponding \str{Fourier}\tr{power} spectra\tr{l density}. Without plasma discharges, the combustor exhibits a strong pulsation at the  frequency of 330 Hz, by looking at the corresponding probability density of the \tr{bandpass} filtered time trace around 330 Hz, it is clear that the system exhibits a \tr{stable} limit cycle\str{ and  thus linearly unstable}. Furthermore, the \str{Fourier}\tr{power} spectra\tr{l density} of the first microphone show a \str{stable}\tr{resonance} peak at 260 Hz. It is worth \str{noting}\tr{mentioning}  that the \tr{thermoacoustic instability} can \tr{alternatively} be \str{stabilized}\tr{suppressed} by adjusting the outlet orifice area at the considered operating point. \str{Therefore, the instability is not due to the intermittent ignition kernels formed in the mixing channel.}

Immediately \tr{after the start of }\str{following} the plasma actuation, \tr{a burst of} the mean \tr{value of the unfiltered dynamic}  pressure signal \str{rises and then decays to the nominal value. The} \tr{is observed and then decay within a few seconds. This burst  is attributed to the rapid change of flame position induced by the start of the NRPD, and a presumed  abrupt variation of the pressure drop across the sequential burner. Nevertheless,  its amplitude and relaxation time do not provide quantitative information about the actual evolution of the mean pressure in the combustor because the} piezoelectric sensor\str{'s} \tr{signal  is }high-pass filter\tr{ed} \str{and}\tr{in} the data acquisition \tr{system}\str{card are responsible for this apparent decay. It is possible that the rise in mean pressure is due to the plasma discharges changing the flow condition around the flame}. \str{However, for t}\tr{T}he purpose of \tr{this work is the study of }\str{studying} thermoacoustic stability \tr{control with NRPD. Therefore, the}\str{, the frequency band near the instability frequency is of interest. The} acoustic signals are bandpass filtered around \tr{the thermoacoustic peak frequency of} 330 Hz and are shown in Figure \ref{fig:micdata}c and  \ref{fig:micdata}d, \str{along}\tr{together} with their corresponding scaled probability density functions (PDFs) in the inset.

As \tr{it can be} seen in figure \ref{fig:micdata}c and \ref{fig:micdata}d, when the plasma is on, the combustor becomes linearly stable. In contrast, without the plasma, the \str{combustor}\tr{PDF of the acoustic pressure} exhibits a bimodal distribution which is a typical feature of a system undergoing a limit cycle. Additionally, the first harmonic at around 660 Hz is also observed in the \str{Fourier}\tr{power} spectra\tr{l density}.
\begin{figure}[t!]
    \centering
    \begin{psfrags}
    \psfrag{aaa}[][]{\scriptsize a)~~}
    \psfrag{bbb}[][]{\scriptsize b)~~}
    \psfrag{ccc}[][]{\scriptsize c)~~}
    \psfrag{ddd}[][]{\scriptsize d)~~}
    \psfrag{eee}[][]{\scriptsize e)~~}
    \psfrag{fff}[][]{\scriptsize f)~~} 
    
    \psfrag{a}[][]{\scriptsize a}
    \psfrag{b}[][]{\scriptsize b}
    \psfrag{c}[][]{\scriptsize c}
    \psfrag{d}[][]{\scriptsize d}
    \psfrag{e}[][]{\scriptsize e}
    \psfrag{f}[][]{\scriptsize f}  
    \psfrag{max}[][]{\scriptsize max}  
    \psfrag{min}[][]{\scriptsize min}  
    
    \psfrag{g}[][]{\scriptsize g)}
    \psfrag{h}[][]{\scriptsize h)}
    \psfrag{intensity}[][]{\scriptsize $\mathrm{I_{OH}}$ (a.u)}
     \psfrag{intensity}[][]{\scriptsize $\mathrm{I_{OH}}$ (a.u)}
    \psfrag{Pressure}[][]{\scriptsize P (Pa)}
    \psfrag{mixingchanqqqqqqqqqq}[][]{\scriptsize Mixing Channel }
    \psfrag{seqcombqqqqqqqqqq}[][]{\scriptsize ~~~Seq. Combustor }
    \psfrag{firstqqqqqqqqqqqqqq}[][]{\scriptsize~ First combustor}
    \psfrag{secondqqqqqqqqqqqqqq}[][]{\scriptsize Seq. combustor~~}

    \psfrag{timems}[][]{\scriptsize{$t$ (ms)}}

    \psfrag{5000}[][]{\scriptsize{5000}}
    \psfrag{-5000}[][]{\scriptsize{-5000}}
    \psfrag{0}[][]{\scriptsize{0}}
    \psfrag{0  }[][]{\scriptsize{0}}
    \psfrag{250}[][]{\scriptsize{250}}
    \psfrag{500}[][]{\scriptsize{500}}
    \psfrag{480}[][]{\scriptsize{480}}
    \psfrag{490}[][]{\scriptsize{490}}
    \psfrag{510}[][]{\scriptsize{510}}
    \psfrag{520}[][]{\scriptsize{520}}
    \psfrag{530}[][]{\scriptsize{530}}
    \psfrag{540}[][]{\scriptsize{540}}
    \psfrag{550}[][]{\scriptsize{550}}
    \psfrag{560}[][]{\scriptsize{560}}
    \includegraphics[trim=0cm 0cm 0cm 0cm,clip,width=1\textwidth]{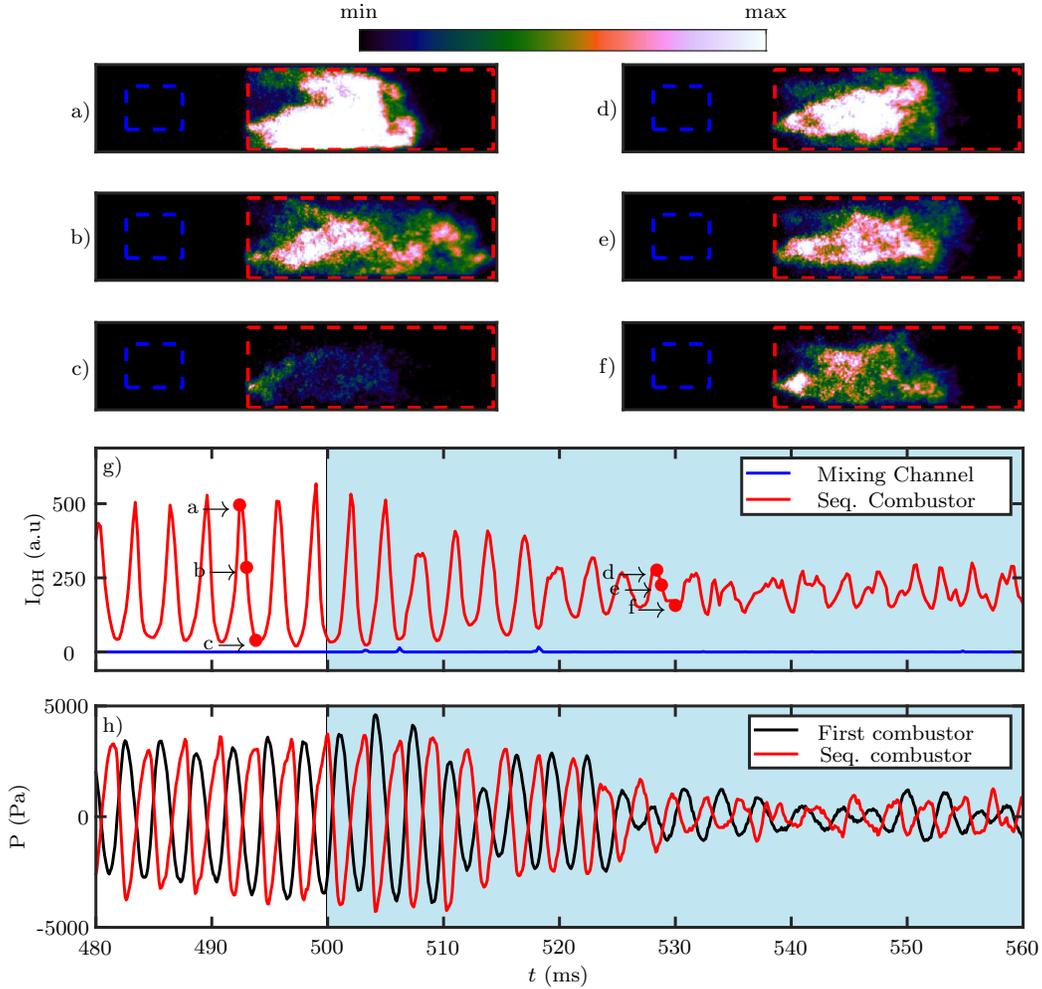}
    \end{psfrags}
    \caption{(a-f) OH chemiluminescence of the sequential combustor at 6 different time instances \bayu{with the time instances shown in g)}. \bayu{The PRF is 10~kHz and the generator voltage is 10~kV. Plasma is initated at $t = 500~ms$}. g) The mean OH intensity inside the mixing channel and sequential combustor. h) The bandpass filtered acoustic pressure signal inside the first and sequential combustor. }
    \label{fig:FlameChem}
\end{figure}

Figures \ref{fig:FlameChem}a to \ref{fig:FlameChem}f \str{display}\tr{show} the OH chemiluminescence of the sequential \str{combustor}\tr{chamber} and \str{a section}\tr{of a portion} of the \tr{burner} mixing channel. In figures \ref{fig:FlameChem}a to \ref{fig:FlameChem}c, the OH chemiluminescence is shown at three different time instances before plasma initiation, while figures \ref{fig:FlameChem}d to \ref{fig:FlameChem}f show the OH chemiluminescence after plasma activation. \bayu{The PRF is at 10~kHz with a generator voltage of 11.7~kV}. The mean intensity within the red and blue squares in figures \ref{fig:FlameChem}a to \ref{fig:FlameChem}f is illustrated in figure \ref{fig:FlameChem}g. Figure \ref{fig:FlameChem}h depicts the acoustic pressure inside the first and sequential combust\str{ors}\tr{ion chambers}. \str{It is noteworthy that prior to plasma actuation, the intensity fluctuates at the same frequency as the acoustic pressure.}Remarkably, the \tr{thermoacoustic limit cycle in the sequential} combustor can be effectively \tr{suppressed} with \tr{a} \bayu{mean} plasma power \tr{of} only 1.1 W, which is about \tr{1.5} $\times~10^{-3}$ \tr{percent} of the thermal power of the flame. \bayu{The mean plasma power in our case is similar to that in \cite{Alkhalifa2022}; however, the thermal power of the flame in our case is 300 times higher than that in their case.}

\begin{figure}[t!]
    \centering
   
    \psfrag{aaa}[][]{\scriptsize a)~~}
    \psfrag{bbb}[][]{\scriptsize b)~~}
    \psfrag{ccc}[][]{\scriptsize c)~~}
    \psfrag{ddd}[][]{\scriptsize d)~~}
    \psfrag{eee}[][]{\scriptsize e)~~}
    \psfrag{fff}[][]{\scriptsize f)~~} 
    
    \psfrag{a}[][]{\scriptsize a}
    \psfrag{b}[][]{\scriptsize b}
    \psfrag{c}[][]{\scriptsize c}
    \psfrag{d}[][]{\scriptsize d}
    \psfrag{e}[][]{\scriptsize e}
    \psfrag{f}[][]{\scriptsize f}    
    \psfrag{g}[][]{\scriptsize g)}
    \psfrag{h}[][]{\scriptsize h)}
    \psfrag{intensity}[][]{\scriptsize $\mathrm{I_{OH}}$ (a.u)}
     \psfrag{intensity}[][]{\scriptsize $\mathrm{I_{OH}}$ (a.u)}
    \psfrag{Pressure}[t][]{\scriptsize $p$ (Pa)}
    \psfrag{FreqHz}[][]{\scriptsize $f$ (Hz)}
    \psfrag{mixingchanqqqqqqqqqq}[][]{\scriptsize Mixing Channel }
    \psfrag{seqcombqqqqqqqqqq}[][]{\scriptsize ~~~Seq. Combustor }
    \psfrag{firstqqqqqqqqqq}[][]{\scriptsize~ First comb.}
    \psfrag{secondqqqqqqqqqq}[][]{\scriptsize Seq. comb. }

    \psfrag{pi2}[][]{\scriptsize $\frac{\pi}{2}$}
    \psfrag{pi3}[][]{\scriptsize $\pi$}
    \psfrag{pi4}[][]{\scriptsize $\frac{3\pi}{2}$}
    \psfrag{pi4}[][]{\scriptsize $\frac{3\pi}{2}$}
    \psfrag{min}[][]{ \scriptsize{min}}
    \psfrag{max}[][]{\scriptsize{max}}
    \psfrag{timems}[][]{\scriptsize{$t$ (ms)}}
    \psfrag{100}[][]{\scriptsize{100}}
    \psfrag{300}[][]{\scriptsize{300}}
    \psfrag{500}[][]{\scriptsize{500}}
    \psfrag{700}[][]{\scriptsize{700}}
    \psfrag{-3000}[][]{\scriptsize{-3000}}
    \psfrag{-1500}[][]{\scriptsize{-1500}}
    \psfrag{1500}[][]{\scriptsize{1500}}
    \psfrag{3000}[][]{\scriptsize{3000}}

    \psfrag{100}[][]{\scriptsize{100}~~}
    \psfrag{120}[][]{\scriptsize{120}~~}
    \psfrag{140}[][]{\scriptsize{140}~~}
    \psfrag{160}[][]{\scriptsize{160}~~}
    \psfrag{180}[][]{\scriptsize{180}~~}

    \psfrag{5000}[][]{\scriptsize{5000}}
    \psfrag{-5000}[][]{\scriptsize{-5000}}
    \psfrag{0}[][]{\scriptsize{0}}
    \psfrag{250}[][]{\scriptsize{250}}
    \psfrag{500}[][]{\scriptsize{500}}

    \psfrag{0.5}[][]{\scriptsize{0.5}~}
    \psfrag{1}[][]{\scriptsize{1~}}
\psfrag{scaled pdf}[l][c][1][270]{\hspace{-5mm} \scriptsize $\hat{P_{p}}$}   
\psfrag{10aaaabbbbcccdddeeefffggg}[][]{\scriptsize{11 kV, $\tr{\eta_p} = 0.5\str{5}\times 10^{-3}\%$}}
    \psfrag{11.7aaaabbbbcccdddeeefffggg}[][]{\scriptsize{11.7 kV, $\tr{\eta_p} = \str{1.4}\tr{1.5}\str{9}\times 10^{-3}\%$}}
    \psfrag{12.5aaaabbbbcccdddeeefffggg}[][]{\scriptsize{12.5 kV, $\tr{\eta_p} = 4.0\str{9}\times 10^{-3}\%$}}
    \psfrag{14aaaabbbbcccdddeeefffggg}[][]{\scriptsize{14 kV, $\tr{\eta_p} = \tr{2.9}\str{29.3}\times 10^{-\str{3}\tr{2}}\%$}}	
    \psfrag{powerdba}[b][]{\scriptsize{$S_{pp}$ (dBa)}}	
    
    \begin{tikzpicture}
       \node[anchor=south west,inner sep=0] at (0,0) {    \includegraphics[trim=0cm 0cm 0cm 0cm,clip,width=1\textwidth]{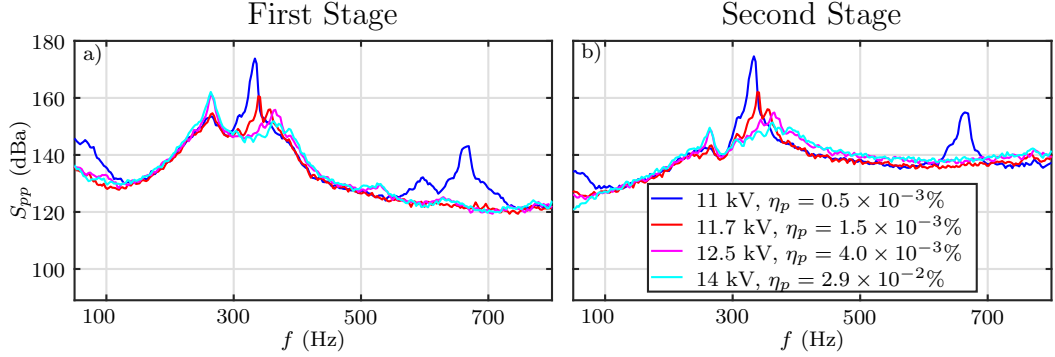}};
       \node[align=center] at (4,4.1) {\bayu{First Stage}\\};
       \node[align=center] at (10.5,4.1) {\bayu{Second Stage}\\};
   \end{tikzpicture}
    
    \caption{The frequency spectra of the acoustic pressure signal with plasma discharges at PRF = 10 kHz and four different generator voltages inside the a) first  and b) second  stage combustor. }
    \label{fig:10kHz_PSD_PDF}
\end{figure}

In Figure \ref{fig:10kHz_PSD_PDF}, the frequency spectra of the \bayu{pressure oscillations} are displayed, with the PRF fixed at 10 kHz and at varying generator voltages. It is evident that the \str{unstable} peak \tr{corresponding to a limit cycle} around 330 Hz becomes smaller\tr{, and thus corresponding to a thermoacoustic stabilization,} and shifts to higher frequencies as the voltage and energy deposition increase. This \tr{frequency} shift can be \tr{indirectly }attributed to a change in the mean flame position, which will be quantified later. In Figure \ref{fig:10kHz_PSD_PDF}a, another peak at around 260 Hz becomes more prominent as the voltage is increased, but \tr{the gaussian-like PDF of the acoustic pressure  filtered around that peak (not shown here), indicates that} it remains \tr{a resonance peak, i.e. the thermoacoustic  oscillations at that frequency are linearly }stable. Notably, the first microphone has a \str{much}more intense peak around 260 Hz compared to the second microphone. However, for the mode at 330 Hz, almost the same amplitudes are observed with both microphones.

\begin{figure}[t!]
    \centering
    \begin{psfrags}
    \psfrag{aaa}[][]{\scriptsize a)~~}
    \psfrag{bbb}[][]{\scriptsize b)~~}
    \psfrag{ccc}[][]{\scriptsize c)~~}
    \psfrag{ddd}[][]{\scriptsize d~~}
    \psfrag{eee}[][]{\scriptsize e~~}
    \psfrag{fff}[][]{\scriptsize f~~} 
    
    \psfrag{Pressure}[t][]{\scriptsize $p$ (Pa)}
    \psfrag{FreqHz}[][]{\scriptsize $f$ (Hz)}
    \psfrag{mixingchanqqqqqqqqqq}[][]{\scriptsize Mixing Channel }
    \psfrag{seqcombqqqqqqqqqq}[][]{\scriptsize ~~~Seq. Combustor }
    \psfrag{firstqqqqqqqqqq}[][]{\scriptsize~ First comb.}
    \psfrag{secondqqqqqqqqqq}[][]{\scriptsize Seq. comb. }

    \psfrag{times}[][]{\scriptsize{$t$ (s)}}
    \psfrag{AmpPa}[][]{\scriptsize{$A$ (Pa)} \vspace{10cm}}
    \psfrag{-2000}[][]{\scriptsize{-2000}~~}
    \psfrag{-4000}[][]{\scriptsize{-4000}~~}
    \psfrag{2000}[][]{\scriptsize{2000}~~}
    \psfrag{4000}[][]{\scriptsize{4000}~~} 

    \psfrag{2000 }[][]{\scriptsize{2000}~~}
    \psfrag{4000 }[][]{\scriptsize{4000}~~} 

    \psfrag{10}[][]{\scriptsize{10}}
    \psfrag{20}[][]{\scriptsize{20}}
    \psfrag{30}[][]{\scriptsize{30}}
    \psfrag{40}[][]{\scriptsize{40}}
    \psfrag{50}[][]{\scriptsize{50}}
    \psfrag{60}[][]{\scriptsize{60}}
    \psfrag{70}[][]{\scriptsize{70}}
    \psfrag{80}[][]{\scriptsize{80}}
    \psfrag{90}[][]{\scriptsize{90}}
    \psfrag{5}[][]{\scriptsize{5}}
    \psfrag{0.05}[][]{\scriptsize{0.05}}
    \psfrag{0.1}[][]{\scriptsize{0.1}}
    \psfrag{0}[][]{\scriptsize{0}}
    \psfrag{0.5}[][]{\scriptsize{~~0.5}}
    \psfrag{0    }[][]{\scriptsize{0}}
    \psfrag{0p}[][]{\scriptsize{0}}
    \psfrag{250}[][]{\scriptsize{250}}
    \psfrag{500}[][]{\scriptsize{500}}
    \psfrag{pressurepa}[][]{\scriptsize{$p$ (Pa)}}
    \psfrag{ProbA}[][]{\scriptsize{$P_A$}}
    \psfrag{pressure}[][]{\scriptsize{$p$ (Pa)}}
        \psfrag{scaledpdf}[][]{\scriptsize{$\hat{P_A}$}}
    \includegraphics[trim=0cm 0cm 0cm 0cm,clip,width=1\textwidth]
    {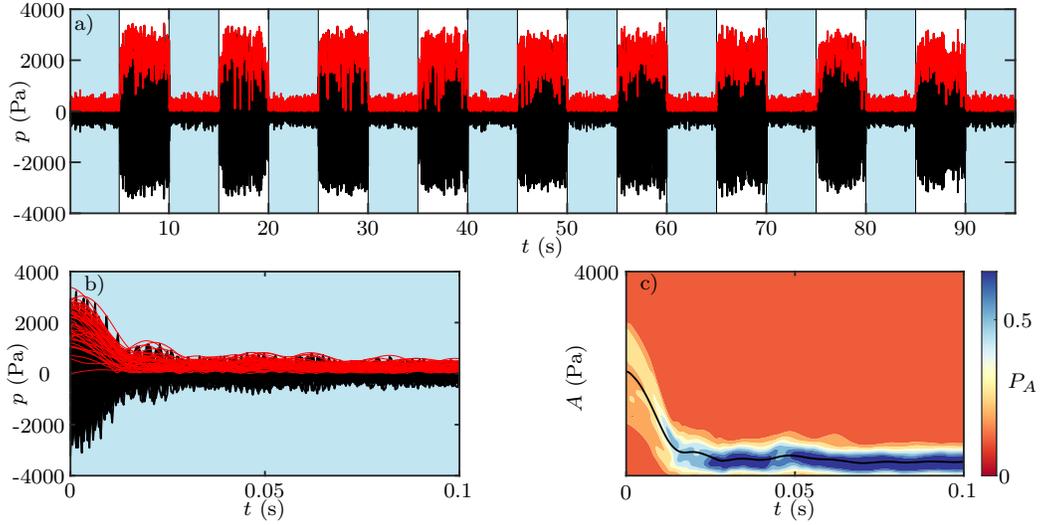}
    \end{psfrags}
    \caption{a) Periodic on-off plasma excitation at PRF = 10 kHz and generator voltage of \bayu{12.5} kV, the modulation frequency is at 0.1 Hz, red and black line depict the envelope and acoustic pressure evolution, respectively. \bayu{The blue shaded regions show the period in which the plasma is on}. b) Cycle to cycle superposition of acoustic pressure time trace. c) The \tr{time evolution of the distribution of the acoustic} pressure envelope \tr{$A$ is shown} \str{distribution} for the first 0.1 second of the actuation. The mean envelope $\tr{\bar{A}}$ is the black line.}
    \label{fig:ControlOnOff}
\end{figure}

\begin{figure}[t!]
    \centering
    \begin{psfrags}

    \psfrag{tms}[][]{\scriptsize{$t$ (ms)}}
    \psfrag{pressure}[][]{\scriptsize{$p_{rms}$ (Pa)}}
 
    \psfrag{2000}[][]{\scriptsize{2000}}
    \psfrag{1800}[][]{\scriptsize{1800}}
    \psfrag{1900}[][]{\scriptsize{1900}}
    \psfrag{50}[][]{\scriptsize{50}}
    \psfrag{100}[][]{\scriptsize{100}}

    \psfrag{0.5}[][]{\scriptsize{0.5}~}
    \psfrag{0.25}[][]{\scriptsize{0.25}~}
    \psfrag{0.75}[][]{\scriptsize{0.75~}}
    \psfrag{1}[][]{\scriptsize{1}~}
    \psfrag{11p7kVaaaa}[][]{\scriptsize{11.7 kV}}
    \psfrag{12p5kVaaaa}[][]{\scriptsize{12.5 kV}}
    \psfrag{14kVaaaa}[][]{\scriptsize{14 kV}}

    \psfrag{0}[][]{\scriptsize{0}}
    \psfrag{0p}[][]{\scriptsize{0}}
    \psfrag{250}[][]{\scriptsize{250}}
    \psfrag{500}[][]{\scriptsize{500}}
    \psfrag{Abarnorm}[][]{\scriptsize{$\hat{A}$ (-)}}
    \includegraphics[trim=0cm 0cm 0cm 0cm,clip,width=0.5\textwidth]
    {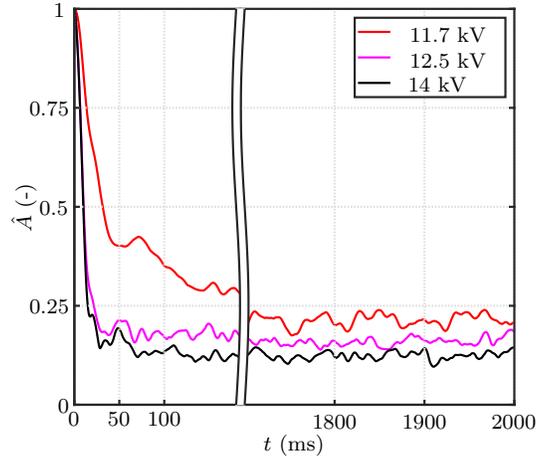}
    \end{psfrags}
    \caption{\tr{Evolution of the normalized m}\str{m}ean \tr{ acoustic pressure} envelope \str{evolution} after plasma is applied at three different generator voltages. The values of the envelope are normalized with respect to their value at $t$ = 0 ms}
    \label{fig:10kHzEnvDecay}
\end{figure}

Another important aspect of the actuator's performance is its ability to stabilize the system quickly, which can be expressed as a decay rate. To measure this \tr{decay}, \tr{in a }similar \tr{way as in} \cite{Noiray2017}, a periodic on-off cycle of plasma actuation is applied. The plasma is turned on for 5 seconds, followed by a 5-seconds off period, and the entire cycle is repeated for 5 minutes. The time trace of the first microphone's bandpass-filtered acoustic signal during this process is shown in Figure~\ref{fig:ControlOnOff}, with the PRF set at 10~kHz and the generator voltage at \bayu{12.5}~kV. The envelope of the acoustic pressure signal \tr{$A$} is obtained from this data \tr{by computing the analytical signal using the Hilbert transform,} and \tr{it is} averaged over the cycles. Figure~ \ref{fig:ControlOnOff} illustrates the distribution of the pressure envelope at different time points, with the black line indicating the mean value. This procedure is repeated at generator voltages of 11.7~kV and \bayu{14}~kV. The resulting decay rates at different voltages are shown in Figure ~\ref{fig:10kHzEnvDecay}, where the envelope is normalized to its value at $t=0$ for better comparison. It is evident that the envelope decays faster as the voltage increases. The system takes around 30 ms to reach a quasi steady-state value with generator voltages of 12.5 and 14 kV, and the steady-state value is lower at higher voltages.

\begin{figure}[t!]
    \centering
    \begin{psfrags}

    \psfrag{tms}[][]{\scriptsize{$t$ (ms)}}
    \psfrag{centreofmass}[][]{\scriptsize{C.O.M}}
    \psfrag{pressure}[][]{\scriptsize{$p_{rms}$ (Pa)}}
 
    \psfrag{20}[][]{\scriptsize{20}~}
    \psfrag{70}[][]{\scriptsize{70}~}
    \psfrag{120}[][]{\scriptsize{120}~}
    \psfrag{170}[][]{\scriptsize{170}~}

    \psfrag{a}[][]{\color{white}\scriptsize{a}~}
    \psfrag{b}[][]{\color{white}\scriptsize{b}~}
    \psfrag{c}[][]{\color{white}\scriptsize{c}~}
    \psfrag{d}[][]{\color{white}\scriptsize{d}~}
    \psfrag{e}[][]{\color{white}\scriptsize{e}}
    \psfrag{f}[][]{\color{white}\scriptsize{f}}
    \psfrag{g}[][]{\color{white}\scriptsize{g}}
    \psfrag{h}[][]{\color{white}\scriptsize{h}}

    \psfrag{min}[][]{\scriptsize{~~min}}
    \psfrag{max}[][]{\scriptsize{~~max}}
    
    \psfrag{distmm}[b][]{\scriptsize{$z$~(mm)}}
    \psfrag{timems}[][]{\vspace{1cm}\scriptsize{$t$~(ms)}}

    \psfrag{0}[][]{\scriptsize{0}}
    \psfrag{520}[][]{\scriptsize{520}}
    \psfrag{480}[][]{\scriptsize{480}}
    \psfrag{500}[][]{\scriptsize{500}}
    \psfrag{Abarnorm}[][]{\scriptsize{$\hat{A}$ (-)}}
    \includegraphics[trim=0cm 0cm 0cm 0cm,clip,width=1\textwidth]
    {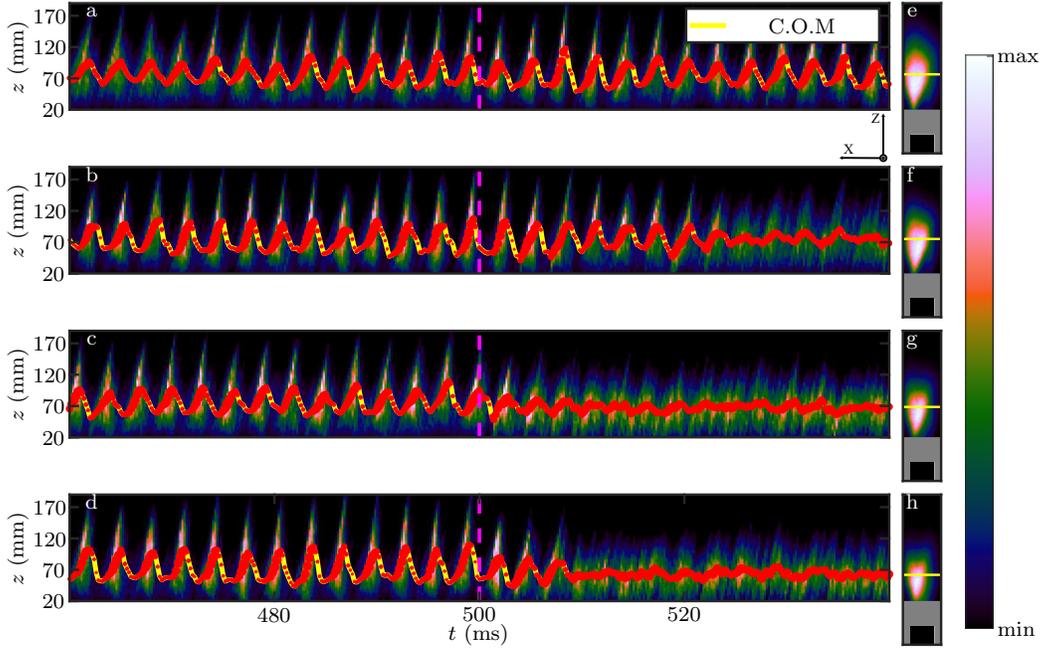}
    \end{psfrags}
    \caption{Evolution of the \bayu{\OHstar{} intensity integrated horizontally along the $x$-axis} and the flame centre of mass before and after the plasma actuation (a-d). \bayu{Only the region in the sequential combustor, which corresponds to the red box in figure \ref{fig:FlameChem}, is considered}. The start of the plasma actuation is at $t$ = 500~ms and indicated by the dashed magenta line in (a-d). The size of the red markers is proportional to the sum of \OHstar{} chemiluminescence intensity. The PRF is fixed at 10~kHz, the generator voltages are a) 11~kV, b) 11.7~kV, c) 12.5~kV, and d) 14~kV. \bayu{The corresponding averaged flame \OHstar{} chemiluminescence images after the plasma initiation for each generator voltage are presented in (e-h)}. }
    \label{fig:10kHzCOM}
\end{figure}

Figure~\ref{fig:10kHzCOM} displays the time evolution of the flame centre of mass at PRF of 10~kHz and various voltages. The \OHstar{} chemiluminescence data are vertically integrated, and the centre of mass is computed along the streamwise direction. At 11 kV, the plasma has little effect on the flame, and the fluctuation around 330 Hz is still evident. With increasing voltage, the fluctuation of the flame centre of mass reduces more rapidly, which is strongly correlated with the pressure signal. Moreover, the steady-state value of the flame centre of mass with plasma actuation decreases as the voltage is increased, and the centre of mass shifts closer to the burner outlet. This shift is due to higher energy deposition, resulting in increased \bayu{mean} plasma power that \str{ignites the mixture} \tr{enhances the autoignition process in the mixing section of the sequential burner } more effectively.

\begin{figure}[t!]
    \centering
    \begin{psfrags}

    \psfrag{COM}[b][]{\color{blue} \scriptsize{COM (mm)}}
    \psfrag{C.O.M}[][]{~~~\scriptsize{C.O.M}}
    \psfrag{NOemission}[][]{\hspace{-0.6cm}\scriptsize{NO}}
    
    \psfrag{rmspressure}[][]{\hspace{-0.6cm}\scriptsize{$p_{rms}$}}
    \psfrag{tms}[][]{\scriptsize{$t$ (ms)}}
    \psfrag{pressure}[][]{\scriptsize{$p_{rms}$ (Pa)}}
    
    \psfrag{noemission}[t][]{\color{red} \scriptsize{$NO~(ppmvd)$}}
    
    \psfrag{200}[][]{\scriptsize{200}~~}
    \psfrag{600}[][]{\scriptsize{600}~~}
    \psfrag{1000}[][]{\scriptsize{1000}~~}
    \psfrag{1400}[][]{\scriptsize{1400}~~}

    \psfrag{64}[][]{\color{blue} \scriptsize{64}~~}
    \psfrag{70}[][]{\color{blue} \scriptsize{70}~~}
    \psfrag{76}[][]{\color{blue} \scriptsize{76}~~}
    \psfrag{82}[][]{\color{blue} \scriptsize{82}~~}

    \psfrag{30}[][]{\color{red} ~~~\scriptsize{8.8}}
    \psfrag{35}[][]{\color{red} ~~~\scriptsize{10.3}}
    \psfrag{40}[][]{\color{red} ~~~\scriptsize{11.7}}
    \psfrag{45}[][]{\color{red} ~~~\scriptsize{13.2}}
    
    \psfrag{11}[][]{\scriptsize{11}}
    \psfrag{11.5}[][]{\scriptsize{11.5}}
    \psfrag{12}[][]{\scriptsize{12}}
    \psfrag{12.5}[][]{\scriptsize{12.5}}
    \psfrag{13}[][]{\scriptsize{13}}
    \psfrag{13.5}[][]{\scriptsize{13.5}}
    \psfrag{14}[][]{\scriptsize{14}}
    \psfrag{NO}[][]{\scriptsize{NO}}
    \psfrag{1}[][]{\scriptsize{1}~}
    \psfrag{11p7kVaaaa}[][]{\scriptsize{11.7 kV}}
    \psfrag{12p5kVaaaa}[][]{\scriptsize{12.5 kV}}
    \psfrag{14kVaaaa}[][]{\scriptsize{14 kV}}

    \psfrag{0}[][]{\scriptsize{0}}
    \psfrag{0p}[][]{\scriptsize{0}}
    \psfrag{250}[][]{\scriptsize{250}}
    \psfrag{500}[][]{\scriptsize{500}}
    \psfrag{AppliedVoltage}[][]{\scriptsize{$V$ (kV)}}
    \includegraphics[trim=0cm 0cm 0cm 0cm,clip,width=0.7\textwidth]
    {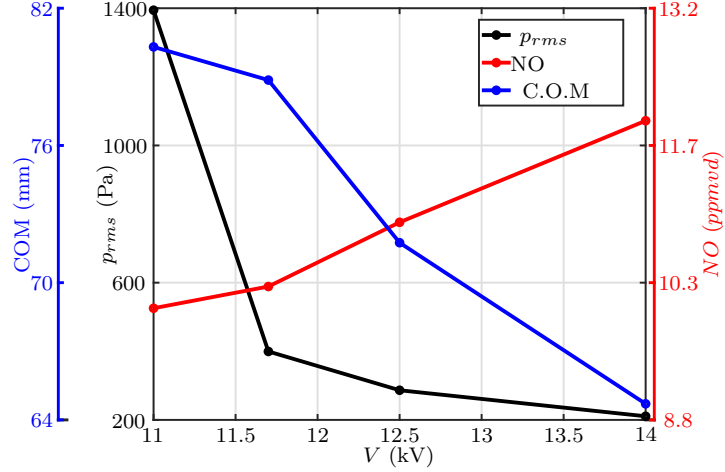}
    \end{psfrags}
    \caption{The dependency of root mean squared of acoustic pressure (band-pass filtered around 330 Hz), NO emission, and the flame centre of mass on the applied generator voltage. The PRF is fixed at 10 kHz.}
    \label{fig:10kHztradeoff}
\end{figure}

Although the decay rate gets faster and \tr{the acoustic} pressure amplitude gets smaller as voltage increases, NO emissions increase slightly. Figure \ref{fig:10kHztradeoff} shows the root mean square of \tr{the acoustic }pressure \tr{$p_{rms}$}, NO emissions, and the flame center of mass with respect to generator voltage. As \tr{it can be }seen, NO emissions increase from around 10~ppmvd at 11 kV to around 12~ppmvd at 14~kV. It is a well-known fact that spark plasma can produce \str{a lot of}\tr{significant} NO \tr{emissions} \cite{Kim2006}. However, in our configuration, the flame center of mass is shifted upstream, resulting in \tr{an} increased residence time for the burnt gases\tr{which can also be the cause of the NO increase}. Furthermore, the energy deposition increases \tr{by an} order of magnitude, from 0.2 mJ to 2 mJ, as depicted in figure \ref{fig:EnergyDepositionvsfreq}. According to \cite{Shcherbanev2022}, the plasma regime changes from glow to spark. However, NO emissions only increase by less than 1~ppmvd. Therefore, the upstream shift of the flame centre of mass is suspected to be the dominant contributor to the increase in NO emissions. The exact mechanism behind this process is not yet clear, and further investigations are needed. 

\begin{figure}[t!]
    \centering
    \begin{psfrags}

    \psfrag{COM}[][]{\color{blue} \scriptsize{COM}}
    \psfrag{C.O.M}[][]{~~~\scriptsize{C.O.M}}
    \psfrag{NOemission}[][]{\hspace{-0.6cm}\scriptsize{NO}}
    
    \psfrag{rmspressure}[][]{\hspace{-0.6cm}\scriptsize{$p_{rms}$}}
    \psfrag{tms}[][]{\scriptsize{$t$ (ms)}}
    \psfrag{pressure}[][]{\scriptsize{$p_{rms}$ (Pa)}}
    
    \psfrag{noemission}[][]{\color{red} \scriptsize{$NO~(mg/mm^3)$}}
    
    \psfrag{10}[][]{\scriptsize{10}}
    \psfrag{20}[][]{\scriptsize{20}}
    \psfrag{40}[][]{\scriptsize{40}}

    \psfrag{11}[][]{\scriptsize{11}~}
    \psfrag{12.5}[][]{\scriptsize{12.5}~}
    \psfrag{14}[][]{\scriptsize{14}~}

    \psfrag{8.8}[][]{\scriptsize{8.8}}
    \psfrag{13.2}[][]{\scriptsize{13.2}}
    \psfrag{17}[][]{\scriptsize{17}}
    
    \psfrag{10}[][]{\scriptsize{10}}
    \psfrag{Voltagekv}[b][]{\scriptsize{$V$~(kV)}}
    \psfrag{PRFkhz}[][]{\scriptsize{$PRF$~(kHz)}}
    \psfrag{12p5kVaaaa}[][]{\scriptsize{12.5 kV}}
    \psfrag{14kVaaaa}[][]{\scriptsize{14 kV}}

    \psfrag{0}[][]{\scriptsize{0}}
    \psfrag{500}[][]{\scriptsize{0}}
    \psfrag{1500}[][]{\scriptsize{~1500}}
    \psfrag{2500}[][]{\scriptsize{~2500}}
    \psfrag{Pa}[l][c][1][0]{\hspace{-0.05cm}\scriptsize{Pa}}
    \psfrag{NO}[l][c][1][0]{\hspace{-0.34cm}\scriptsize{ppmvd}}    
    \psfrag{AppliedVoltage}[][]{\scriptsize{$V$ (kV)}}
    \includegraphics[trim=0cm 0cm 0cm 0cm,clip,width=1\textwidth]
    {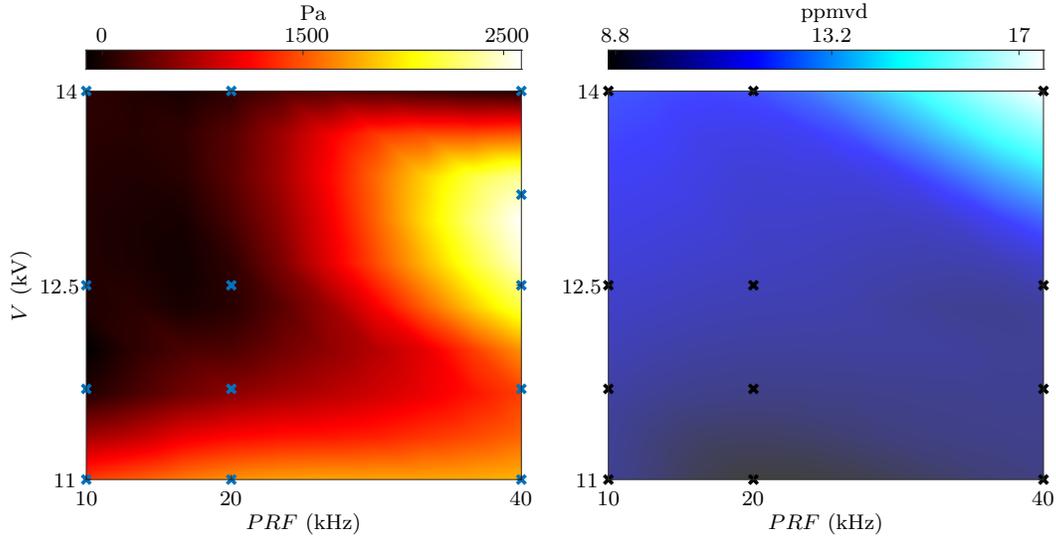}
    \end{psfrags}
    \caption{ \str{Surface}\tr{Contour}  map\tr{s} of the rms acoustic pressure  (left) and the NO emission (right) with respect to the variation of plasma repetition frequency and \str{generator} \tr{pulses} voltage. The crosses indicate the measurement points. The bandwidth for the rms \tr{acoustic} pressure calculation is from 200 Hz to 400 Hz. The NO measurement at V = 13.2 kV and PRF = 40 kHz is not available \str{due to high pulsation of the combustor}\tr{because the combustor cannot be operated for a long enough time interval at this condition due to excessive acoustic amplitudes}.}
    \label{fig:rmsmap}
\end{figure}

The same \tr{approach was followed}\str{routines were performed} at higher PRFs, and the resulting\str{ root mean squared (rms) pressure} \tr{$p_{rms}$} and NO emission map\tr{s} are shown in figure \ref{fig:rmsmap}. Since the\str{re is a} \tr{thermoacoustic }peak \tr{of interest in the power spectral density of the acoustic pressure is} at around 260 Hz\str{ that needs to be considered}, the bandpass filter was set to span from 200 Hz to 400 Hz for the rms calculation. Examining the map, we observe that at 11.7 kV, the plasma discharges at PRF = 10 and 20 kHz are more effective than that at PRF = 40 kHz. This is consistent with the \tr{fact that the }\bayu{mean} plasma power \tr{obtained at 40 kHz for this voltage is lower than the one at 10 kHz or 20 kHz} as shown in figure \ref{fig:EnergyDepositionvsfreq}b. At \str{this condition}\tr{40 kHz and 11.7 kV}, the plasma is \tr{thus} not strong enough to affect the system. However, at 12.5 kV, all \bayu{mean} plasma powers at all PRFs are in the same order of magnitude \tr{(see figure \ref{fig:EnergyDepositionvsfreq}b)}. When the PRF is set to 40 kHz, the rms pressure increases. The same behavior is observed at 13.2 kV, but the system is stabilized again when the voltage is set to 14 kV. Because there is a strong pulsation at V = 13.2 kV and PRF = 40 kHz, NO measurement could not be done. Nevertheless, it is clear that the NO emission map shows an increasing trends towards high PRF and high generator voltage\tr{, which is consistent with the findings in \cite{Xiong2019}, which investigated the effect of plasma on the sequential flame position and on the NO emissions in another sequential combustor}. By looking at the \str{surface}\tr{contour} maps, it is evident that staying at PRF = 10 kHz \str{and generator}\tr{with a pulse}  voltage above 11 kV, the \tr{thermoacoustic eigenmode in the} combustor can be effectively stabilized without compromising the NO emissions. The dependency of NO emissions on the flame centre of mass and \bayu{mean} plasma power for all operating points will be further discussed in the subsequent paragraphs.

\begin{figure}[t!]
    \centering
    
    \psfrag{aaa}[][]{\scriptsize a)}
    \psfrag{bbb}[][]{\scriptsize b)~~}
    \psfrag{ccc}[][]{\scriptsize ~c)}
    \psfrag{ddd}[][]{\scriptsize ~d)}
    \psfrag{eee}[][]{\scriptsize ~e)}
    \psfrag{fff}[][]{\scriptsize ~~f)} 
    
    \psfrag{a}[][]{\scriptsize a}
    \psfrag{b}[][]{\scriptsize b}
    \psfrag{c}[][]{\scriptsize c}
    \psfrag{d}[][]{\scriptsize d}
    \psfrag{e}[][]{\scriptsize e}
    \psfrag{f}[][]{\scriptsize f}    
    \psfrag{g}[][]{\scriptsize g}
    \psfrag{h}[][]{\scriptsize h}
    \psfrag{intensity}[][]{\scriptsize $\mathrm{I_{OH}}$ (a.u)}
     \psfrag{intensity}[][]{\scriptsize $\mathrm{I_{OH}}$ (a.u)}
    \psfrag{Pressure}[t][]{\scriptsize $p$ (Pa)}
    \psfrag{FreqHz}[][]{\scriptsize $f$ (Hz)}
    \psfrag{mixingchanqqqqqqqqqq}[][]{\scriptsize Mixing Channel }
    \psfrag{seqcombqqqqqqqqqq}[][]{\scriptsize ~~~Seq. Combustor }
    \psfrag{firstqqqqqqqqqq}[][]{\scriptsize~ First comb.}
    \psfrag{secondqqqqqqqqqq}[][]{\scriptsize Seq. comb. }

    \psfrag{pi2}[][]{\scriptsize $\frac{\pi}{2}$}
    \psfrag{pi3}[][]{\scriptsize $\pi$}
    \psfrag{pi4}[][]{\scriptsize $\frac{3\pi}{2}$}
    \psfrag{pi4}[][]{\scriptsize $\frac{3\pi}{2}$}
    \psfrag{min}[][]{ \scriptsize{min}}
    \psfrag{max}[][]{\scriptsize{max}}
    \psfrag{timems}[][]{\scriptsize{$t$ (ms)}}
    \psfrag{100}[][]{\scriptsize{100}}
    \psfrag{300}[][]{\scriptsize{300}}
    \psfrag{500}[][]{\scriptsize{500}}
    \psfrag{700}[][]{\scriptsize{700}}
    \psfrag{-6000}[][]{\scriptsize{~-6000}}
    \psfrag{-1500}[][]{\scriptsize{-1500}}
    \psfrag{1500}[][]{\scriptsize{1500}}
    \psfrag{6000}[][]{\scriptsize{6000}}

    \psfrag{100}[][]{\scriptsize{100}~~}
    \psfrag{120}[][]{\scriptsize{120}~~}
    \psfrag{140}[][]{\scriptsize{140}~~}
    \psfrag{160}[][]{\scriptsize{160}~~}
    \psfrag{180}[][]{\scriptsize{180}~~}

    \psfrag{5000}[][]{\scriptsize{5000}}
    \psfrag{-5000}[][]{\scriptsize{-5000}}
    \psfrag{0}[][]{\scriptsize{0}}
    \psfrag{250}[][]{\scriptsize{250}}
    \psfrag{500}[][]{\scriptsize{500}}

    \psfrag{0.5}[][]{\scriptsize{0.5}~}
    \psfrag{1}[][]{\scriptsize{1~}}
\psfrag{scaled pdf}[l][c][1][270]{\hspace{-5mm} \scriptsize $\hat{P_{p}}$}   
\psfrag{10aaaa}[][]{\scriptsize{11 kV}}
    \psfrag{11.7aaaa}[][]{\scriptsize{11.7 kV}}
    \psfrag{12.5aaaa}[][]{\scriptsize{12.5 kV}}
    \psfrag{13.2aaaa}[][]{\scriptsize{13.2 kV}}
    \psfrag{14aaaa}[][]{\scriptsize{14 kV}}	
    \psfrag{powerdba}[b][]{\scriptsize{$S_{pp}$ (dBa)}}	

       \begin{tikzpicture}
       \node[anchor=south west,inner sep=0] at (0,0) {    \includegraphics[trim=0cm 0cm 0cm 0cm,clip,width=1\textwidth]{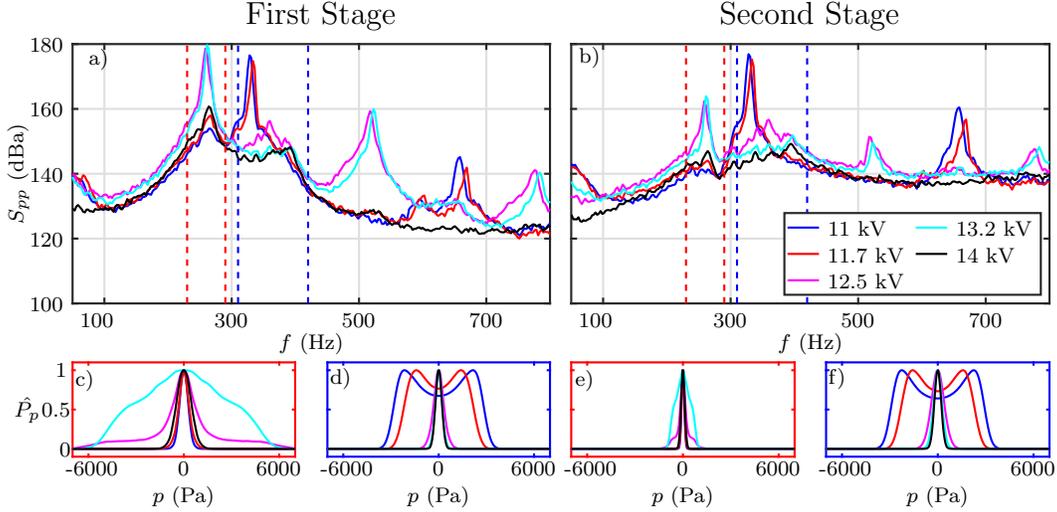}};
       \node[align=center] at (4,6) {\bayu{First Stage}\\};
       \node[align=center] at (10.5,6) {\bayu{Second Stage}\\};
   \end{tikzpicture}

    \caption{The frequency spectra of the acoustic pressure signal with plasma discharges at $\mathrm{PRF} = \mathrm{40~kHz}$ at four different generator voltages inside the a) first  and b) second  stage combustor. The acoustic presssure histogram of the bandpass filtered signal inside the first stage combustor at around 260 Hz c) and 330 Hz d). The acoustic presssure histogram of the bandpass filtered signal inside the second stage combustor at around 260 Hz e) and 330 Hz f). Another mode at around 260 Hz is excited when the generator voltage is at 12.5 kV and 13.2 kV.}
    \label{fig:40kHz_PSD_PDF}
\end{figure}

To shed light on the peculiar phenomenon at PRF = 40 kHz, \str{we plotted the frequency spectra}\tr{it is interesting to show the power spectral density}  of both microphones\tr{. This information is given} in figure \ref{fig:40kHz_PSD_PDF}. \str{As shown,}\tr{Notably,} at 12.5 kV and 13.2 kV, the mode at 260 Hz  \str{is excited}\tr{becomes self-excited and exhibits a very high amplitude of 180 dBa}, while the mode at 330 Hz is stabilized. The \str{distribution} \tr{PDF of the acoustic pressure $\hat{P_p}$ shown in figure \ref{fig:40kHz_PSD_PDF}c} exhibit \tr{in the case of repetitive pulses of 13.2 kV} a typical feature of intermittently unstable \str{flames}\tr{thermoacoustic system}\str{, as seen in the pressure density plot in figure \ref{fig:40kHz_PSD_PDF} c)}. This observation is also confirmed by the time trace of the filtered signal. According to \cite{Bonciolini2021}, \str{this}\tr{such} intermittent behavior can be \str{attributed to the colored-driven}\tr{caused by random fluctuations of the} time delay \str{perturbation} of the flame response \tr{to acoustic perturbations around the mean time delay}. \tr{For time delay fluctuations that can be described by an Ornstein-Uhlenbeck process, intermittent high amplitude bursts of oscillations occur when these fluctuations induce excursions of the system in linearly unstable conditions, and when the correlation time of the fluctuations is long enough to allow the thermoacoustic system to adapt to the random changes of stability \cite{Bonciolini2021}.} In \str{our}\tr{the present} case, the \tr{fluctuating time history of the} ignition kernels produced by the plasma could be the source of this time delay perturbation. However, to identify the exact reason for this behavior, further thermoacoustic analysis is required, which will be the subject of future investigations.

\begin{figure}[t!]
    \centering
    \begin{psfrags}
    \psfrag{aaa}[][]{\scriptsize a)~~}
    \psfrag{bbb}[][]{\scriptsize b)~~}
    \psfrag{ccc}[][]{\scriptsize c)~~}
    \psfrag{ddd}[][]{\scriptsize d)~~}
    \psfrag{eee}[][]{\scriptsize e)~~}
    \psfrag{fff}[][]{\scriptsize f)~~} 
    
    \psfrag{a}[][]{\textbf{\scriptsize a}}
    \psfrag{b}[][]{\textbf{\scriptsize b}}
    \psfrag{c}[][]{\textbf{\scriptsize c}}
    \psfrag{d}[][]{\textbf{\scriptsize d}}
    \psfrag{e}[][]{\textbf{\scriptsize e}}
    \psfrag{f}[][]{\textbf{\scriptsize f}}    
    \psfrag{g}[][]{\scriptsize g)}
    \psfrag{h}[][]{\scriptsize h)}
    \psfrag{intensity}[][]{\scriptsize $\mathrm{I_{OH}}$ (a.u)}
     \psfrag{intensity}[][]{\scriptsize $\mathrm{I_{OH}}$ (a.u)}
    \psfrag{Pressure}[][]{\scriptsize P (Pa)}
    \psfrag{mixingchanqqqqqqqqqq}[][]{\scriptsize Mixing Channel }
    \psfrag{seqcombqqqqqqqqqq}[][]{\scriptsize ~~~Seq. Combustor }
    \psfrag{firstqqqqqqqqqqqqqq}[][]{\scriptsize~ First combustor}
    \psfrag{secondqqqqqqqqqqqqqq}[][]{\scriptsize Seq. combustor~~ }

    \psfrag{timems}[][]{\scriptsize{$t$ (ms)}}

    \psfrag{5000}[][]{\scriptsize{5000}}
    \psfrag{-5000}[][]{\scriptsize{-5000}}
    \psfrag{0}[][]{\scriptsize{0}}
    \psfrag{0  }[][]{\scriptsize{0}}
    \psfrag{250}[][]{\scriptsize{250}}
    \psfrag{500}[][]{\scriptsize{500}}
    \psfrag{480}[][]{\scriptsize{480}}
    \psfrag{490}[][]{\scriptsize{490}}
    \psfrag{510}[][]{\scriptsize{510}}
    \psfrag{520}[][]{\scriptsize{520}}
    \psfrag{530}[][]{\scriptsize{530}}
    \psfrag{540}[][]{\scriptsize{540}}
    \psfrag{550}[][]{\scriptsize{550}}
    \psfrag{560}[][]{\scriptsize{560}}
    \psfrag{580}[][]{\scriptsize{580}}
    \psfrag{600}[][]{\scriptsize{600}}
    \psfrag{620}[][]{\scriptsize{620}}
    \psfrag{640}[][]{\scriptsize{640}}
    \psfrag{660}[][]{\scriptsize{660}}

        \psfrag{max}[][]{\scriptsize max}  
    \psfrag{min}[][]{\scriptsize min} 
    
    \includegraphics[trim=0cm 0cm 0cm 0cm,clip,width=1\textwidth]{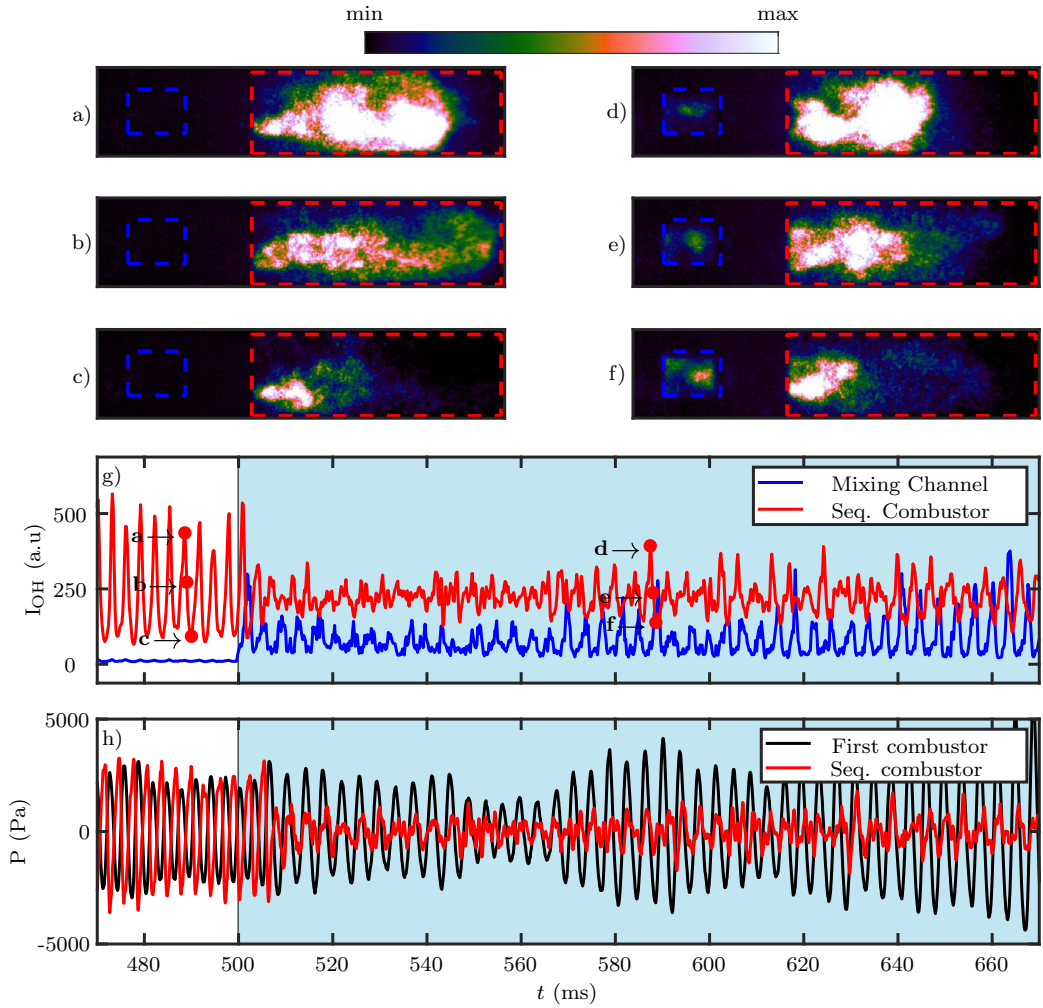}
    \end{psfrags}
    \caption{(a-f) OH chemiluminescence of the sequential combustor at 6 different time instances, the PRF is at 40 kHz and the generator voltage is at 13.2 kV. g) the mean OH intensity inside the mixing channel and sequential combustor. h) The bandpass filtered acoustic pressure signal inside the first and sequential combustor.}
    \label{fig:FlameChem40kHz}
\end{figure}

The OH chemiluminescence signals at 13.2 kV and PRF = 40 kHz, recorded \str{at different time instances} \tr{during a time interval when the NRPD was activated}, are shown in figures \ref{fig:FlameChem40kHz}a to \ref{fig:FlameChem40kHz}f. \tr{When the NRPD actuation is turned on with these pulse generator settings, the thermoacoustic dynamics changes from a robust limit cycle at 330 Hz, to a limit cycle at 260 Hz, and in contrast with the PRF of 10 kHz and  20 kHz, the thermoacoustic system is not stabilized.} As \tr{it can be} seen \tr{in figure \ref{fig:FlameChem40kHz}}, OH chemiluminescence \str{signals are observed} \tr{is visible} inside the mixing channel after plasma actuation, due to the formation of \tr{ignition} kernels induced by the \str{plasma}\tr{NRPD}. \str{The time trace of the mean OH intensity in f}\tr{F}igure \ref{fig:FlameChem40kHz}g shows \str{that} the OH \tr{chemiluminescence} intensity inside the mixing channel\tr{, which} fluctuates \str{along with that inside} \tr{as the one in} the combustion chamber at \tr{the acoustic pressure oscillation} \str{a} frequency of 260 Hz\str{, which is also the frequency of the pressure pulsation}. The \tr{acoustic} pressure time trace \str{depicted}\tr{shown} in figure \ref{fig:FlameChem40kHz}h clearly \str{shows}\tr{indicates} that, prior to plasma actuation, both microphones record\str{ed} similar \tr{acoustic} pressure amplitudes. However, after the \str{plasma discharges were}\tr{NRPD are} applied, the \tr{acoustic} pressure \str{pulsation} in the first stage combustor \str{was much}\tr{is significantly}  higher than that in the sequential combustor. This observation is consistent with figure \ref{fig:40kHz_PSD_PDF}, which shows a 20 dB difference in power spectral density at 260 Hz between the two microphones. It appears that the \tr{thermoacoustic} mode at 260 Hz is more localized inside the first combustor than in the second combustor. \str{However further investigations using a Helmholtz solver or thermoacoustic network model will be needed to study the modes of the combustor.}

\begin{figure}[t!]
    \centering
    \begin{psfrags}

    \psfrag{pressurepa}[][]{\scriptsize $p$ (Pa)}
    \psfrag{times}[][]{\scriptsize $t$ (s)}
    \psfrag{WithPlasma}[][]{\scriptsize{with plasma}}
    \psfrag{WithoutPlasmaxxx}[][]{\hspace{-0.4cm}\scriptsize{without plasma}}    
    \psfrag{6000}[][]{\scriptsize{6000}}    
    \psfrag{3000}[][]{\scriptsize{3000}}
    \psfrag{0}[][]{\scriptsize{0}}
    \psfrag{0.01}[][]{\scriptsize{0.01}}
    \psfrag{0.02}[][]{\scriptsize{0.02}}
    \psfrag{0.03}[][]{\scriptsize{0.03}}
    \psfrag{0.04}[][]{\scriptsize{0.04}}
    \psfrag{0.05}[][]{\scriptsize{0.05}}
    
    \psfrag{-3000}[][]{\scriptsize{-3000}}
    \psfrag{-6000}[][]{\scriptsize{-6000}}
    \psfrag{80}[][]{\scriptsize{80}}    
    \psfrag{2mm}[][]{\color{white}\scriptsize{2 mm}}

    \includegraphics[trim=0cm 0cm 0cm 0cm,clip,width=1\textwidth]
    {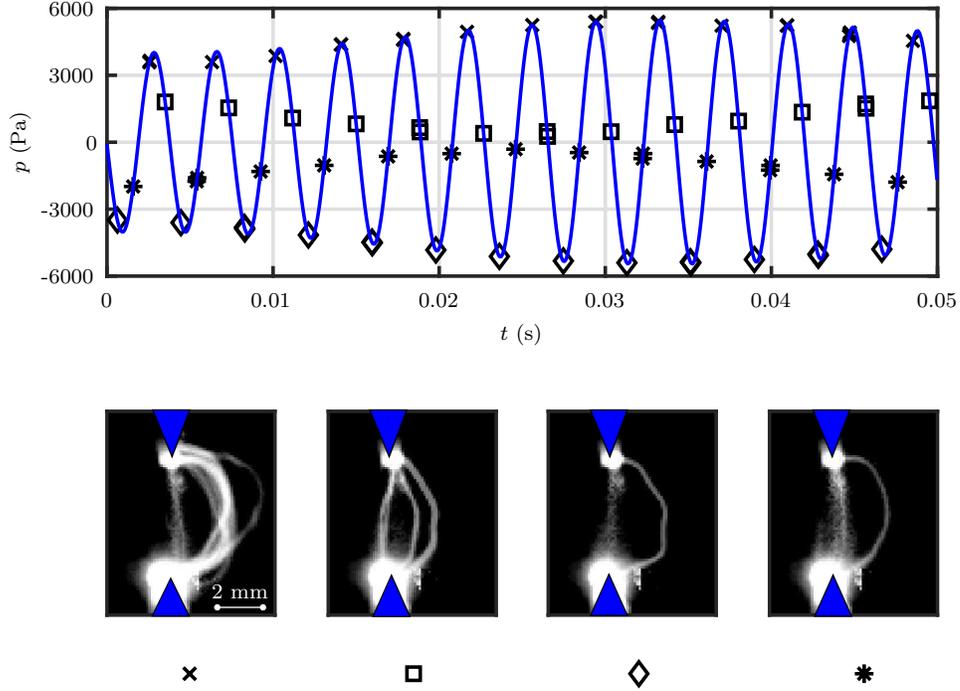}
    \end{psfrags}
    \caption{Phase average of the plasma discharge at $V$ = 13.2~kV and PRF = 40~kHz (second row). The reference signal for the phase averaging is the microphone inside the first stage combustion chamber (first row).}
    \label{fig:DischargePhaseAvg}
\end{figure}

The plasma discharges are visualized in figure \ref{fig:DischargePhaseAvg}. The images are phase averaged with respect to the bandpass-filtered acoustic signals of the first microphone. As \tr{it} can be seen, \tr{during an oscillation cycle,} when the \tr{acoustic} pressure \tr{in} the first stage combustor reaches the maximum point, the plasma bends \str{more}  towards the outlet of the sequential burner. Whereas at the other phase angles, the discharge channels are relatively straight. The plasma bending effect is similar to the one observed in \cite{Shcherbanev2022}. \tr{In this reference}, the bending \bayu{occurs} because the inter-pulse time of the discharges is close to the convective time\tr{, but in contrast to the present work, t}here was no thermoacoustic instability \tr{at the considered operating conditions}. In  \tr{the present work}, the thermoacoustic instability leads to the synchronization of the periodic plasma channel bending with the acoustic \str{signal}\tr{field}.

\begin{figure}[t!]
    \centering
    \begin{psfrags}

    \psfrag{NOemission}[b][]{\hspace{-0.6cm}\scriptsize{NO~($\mathrm{ppmvd}$)}}
    \psfrag{COMmm}[][]{\scriptsize{COM (mm)}}
    
    \psfrag{WithPlasma}[][]{\scriptsize{with plasma}}
    \psfrag{WithoutPlasmaxxx}[][]{\hspace{-0.4cm}\scriptsize{without plasma}}    
    \psfrag{20}[][]{\scriptsize{5.88}~~}   
    \psfrag{a}[][]{\scriptsize{a)}} 
    \psfrag{b}[][]{\scriptsize{b)}}  
    \psfrag{30}[][]{\scriptsize{8.82}~~}
    \psfrag{40}[][]{\scriptsize{11.76}~~}
    \psfrag{50}[][]{\scriptsize{14.70}~~}
    \psfrag{60}[][]{\scriptsize{17.65}~~}
    \psfrag{360}[][]{\scriptsize{360}}    
    \psfrag{380}[][]{\scriptsize{380}}
    \psfrag{400}[][]{\scriptsize{400}}
    \psfrag{1e-1}[][]{\scriptsize{$10^{-1}$}}    
    \psfrag{1e0}[][]{\scriptsize{1}}
    \psfrag{1e1}[][]{\scriptsize{$10^1$}}
    \psfrag{1e2}[][]{\scriptsize{$10^2$}}
    \psfrag{PlasmaPower}[][]{\scriptsize{P (W)}}
    \psfrag{reldistprobe}[][]{\scriptsize{$x_p - COM$ (mm)}}
    
    \includegraphics[trim=0cm 0cm 0cm 0cm,clip,width=1\textwidth]
    {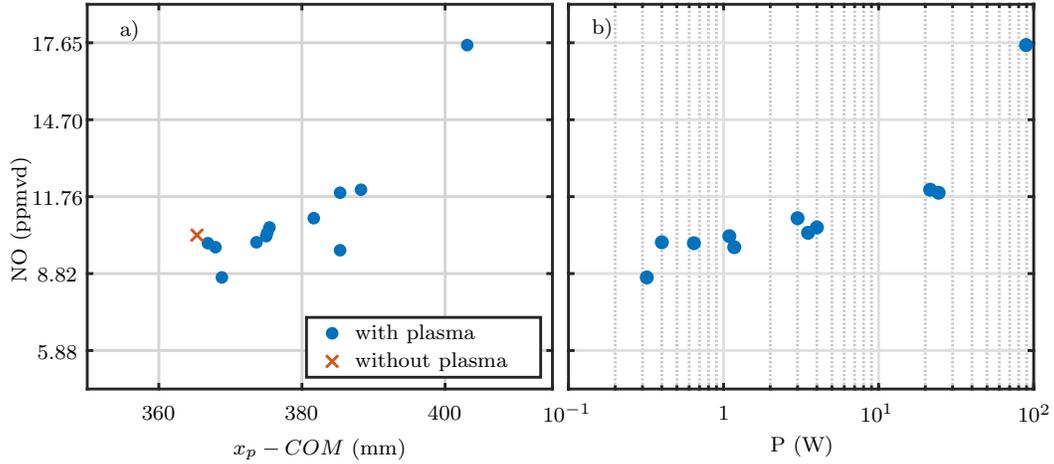}
    \end{psfrags}
    \caption{a) NO emissions as a function of the relative distance between the probe location and the flame centre of mass. \bayu{The red cross is the NO measurement at thermoacoustically stable condition without plasma, obtained through adjusting the outlet orifice geometry.} b) NO emissions as a function of the \bayu{mean} plasma power.}
    \label{fig:COMvsNOx}
\end{figure}

Figure \ref{fig:COMvsNOx}a \str{displays}\tr{shows} the NO emissions of the combustor with \str{plasma operated at} different \tr{NRPD} voltages and PRF, plotted against the relative distance between the emission probe location and the flame centre of mass. \bayu{One measurement point without the plasma, which is indicated by a red cross in figure \ref{fig:COMvsNOx}a, is obtained by stabilizing the combustor through adjusting the outlet orifice.} The maximum \tr{NO} emission of approximately 17.65~ppmvd occurs at a generator voltage of 14~kV and a PRF of 40~kHz, with a general increasing trend observed as the relative distance increases. It is worth noting that high PRF and voltage can cause \str{some of the flames to penetrate into} \tr{early ignition inside} the \tr{sequential burner} mixing channel, leading to a reduction in mixing quality between the vitiated flow and \tr{the} secondary fuel and \tr{consequently a potential}\str{an} increase in NO emissions. In Figure \ref{fig:COMvsNOx}b, a weak correlation between NO emissions and \bayu{mean} plasma power is observed at power levels ranging from 0.1 to 10~W. However, as the \bayu{mean} plasma power increases to 20-81~W, a positive correlation is observed, potentially due to the residence time of the burnt gases or plasma-generated NO. \str{However, if}\tr{Still, as}  the  goal \tr{of the NRPD actuation in this work} is to thermoacoustically stabilize the \str{flame} \tr{sequential combustor}, a generator voltage of 11.7~kV and PRF of 10~kHz \str{can achieve satisfactory results}\tr{is obviously the optimum because the plasma does not lead to an increase of the NO emissions compared to the non-actuated operation.}\str{, with comparable NO emissions to those without plasma.}
\str{T}\tr{Furthermore, t}he \tr{present ultra-low-power NRPD-based} control strateg\tr{ies}\tr{y for sequential combustors} \str{can be  improved} \tr{opens other possibilities such as thermoacoustic instability control during transient operation } by employing a feedback \tr{loop control.} \str{based strategy similar to \cite{Moeck2013}. With this strategy, the duty cycle of the plasma can be reduced to about 50 $\%$ yielding to a lower \bayu{mean} plasma power and possibly lower NO emissions. }


\section{Conclusions and outlook}

\str{The measurement series have }\tr{This study} demonstrate\str{d}\tr{s} that \tr{ultra-low-power non-equilibrium} plasma \str{discharges can serve as an} \tr{can} effective\tr{ly} \str{actuator to} stabilize a thermoacoustically unstable sequential combustor\tr{. Indeed, the power of the plasma produced by the NRPD which can achieve thermoacoustic stabilization, can be  5 orders of magnitude} \str{with very} low\tr{er} \str{power} compared to the flame thermal power\str{. W}\tr{: w}ith a \bayu{mean} plasma power of 1.1~W, which \str{amounts to} \tr{is} $\str{1.4}\tr{1.5}\times10^{-3}\str{\%}$ \tr{percent} of the flame thermal power \tr{of 73.4 kW}, the \tr{thermoacoustic limit cycle in the} sequential combustor was successfully stabilized. At this condition, there is practically no additional NO emission compared to the situation where the \str{flame} \tr{thermoaocustic mode} was stabilized by changing the outlet orifice \tr{geometry} of the combustor \tr{with a motor-driven water-cooled piston}. However, at PRF = 40~kHz and generator voltages of 12.5~kV, and 13.2~kV, another \tr{thermoaocustic} mode at 260 Hz \str{was excited}\tr{becomes self-excited}. Therefore, \str{a careful and thorough investigation or a} \tr{now that an effective ultra-low-power actuator has been found for sequential combustors in the form of NRPD in the sequential burner, a significant part of the research efforts should concentrate on the development of feedback control for ensuring optimum trade-off between global thermoacoustic stability and NO emissions during stead and transient operation. } \str{safe optimization procedure is required to implement it in real systems. Furthermore, a significant increase in NO emissions might occur at high generator voltage and PRF. Hence, it is essential to weigh the trade-offs. F}\tr{Finally, f}urther investigations are required to \str{understand the stabilization mechanism as well as mode switching observed in the experiment. Finally, operating} \tr{demonstrate the practicality of this NRPD actuation} at elevated pressures\str{ will be the next crucial step to evaluate the applicability of NRPD in real systems}\tr{, and to develop predicting tools of the interaction between plasma kinetics and combustion reactions in the thermochemical environment of the turbulent sequential burner}.

\section*{Acknowledgements}
This project has received funding from the European Research Council (ERC) under the European Union’s Horizon 2020 research and innovation program (grant agreement No [820091]).

\bibliographystyle{elsarticle-num}
\bibliography{cas-refs}

\end{document}